\documentclass{aa}
\usepackage{graphicx}
\usepackage{txfonts}
\usepackage{amssymb}

\begin{document}

\title{The T\,Tauri star RY\,Tau as a case study of the inner regions of circumstellar dust disks}
\author{A. A. Schegerer \inst{1} \and S. Wolf \inst{1} \and Th. Ratzka \inst{2} \and 
Ch. Leinert \inst{1}}

\offprints{A. A. Schegerer, \email{schegerer@mpia-hd.mpg.de}}

\institute{Max-Planck-Institut f\"ur Astronomie, K\"onigstuhl 17, 69117 Heidelberg, Germany \and
Astrophysikalisches Institut Potsdam, An der Sternwarte 16, 14482 Potsdam, Germany}

\date{Received $<>$ ; accepted $<>$ }

\abstract
{}
{We study the inner region ($\sim$$1.0\ \mathrm{AU}$ up to a few $10\ \mathrm{AUs}$) of the 
  circumstellar disk around the ``classical'' T\,Tauri star \object{RY\,Tau}. Our aim is to find a 
  physical description satisfying the available interferometric data, obtained with the 
  mid-infrared interferometric instrument at the Very Large Telescope Interferometer,   
  as well as the spectral energy distribution in the visible to millimeter wavelength 
  range. We also compare the findings with the results of similar studies,
  including those of intermediate-mass Herbig Ae/Be stars.
}
{Our analysis is done within the framework of a passively heated circumstellar disk, which is 
  optionally supplemented by the effects of accretion and an added envelope. To achieve a more 
  consistent and realistic model, we used our continuum transfer code MC3D. In addition, we studied 
  the shape of the $10\ \mathrm{\mu m}$ silicate emission feature in terms of the underlying dust 
  population, both for single-dish and for interferometric measurements. 
}
{We show that a modestly flaring disk model with accretion can explain both the observed 
  spectral energy distribution and the mid-infrared visibilities obtained with the mid-infrared 
  infrared instrument. 
  We found an interesting ambiguity: a circumstellar active disk model with an added
  envelope, and a lower accretion rate than in the active disk model without envelope,
  could represent the observations equally as well. This type of model with
  the envelope should be considered a viable 
  alternative in future models of other T\,Tauri stars. The approach of a disk
  with a puffed-up inner rim wall and the influence of a stellar companion is also 
  discussed. We also investigate the
  influence of various fit parameters on the outcome of the radiative 
  transfer modeling. From the study of the silicate emission feature we see evidence for dust 
  evolution in a T\,Tauri star, with a decreasing fraction of small amorphous and an increasing 
  fraction of crystalline particles closer to the star.
}
{}


\authorrunning{Schegerer et al.}
\titlerunning{The inner regions of circumstellar dust disks}
\maketitle

\section{Introduction}\label{section:introduction}
T\,Tauri stars are known as precursors of low-mass main sequence stars ($\lse 2-3\ 
\mathrm{M_{\odot}}$). In contrast to main sequence stars, their characteristic properties are strong 
emission line radiation (e.\/g., Balmer $\alpha$) and excessive continuum
radiation observed in the UV, infrared 
and the millimeter (mm) wavelength range of their spectral energy distribution (SED). It has 
been shown that the spatial 
distribution of circumstellar dust in a disk or an envelope that is primarily exposed to 
stellar radiation, is responsible for the excess radiation in the infrared wavelength range (e.\/g., 
Adams et al.~\cite{adams}), while accretion of circumstellar material results in the UV excess and 
strong emission line radiation (see Hartmann~\cite{hartmannIII} for a review).

The extraordinary interest in the inner region of a circumstellar disk results from the assumption 
that the formation of planets is favored there (see Nagasawa et al.~\cite{nagasawa}; 
W\"unsch et al.~\cite{wuensch}; Klahr~\cite{klahr}). While mm observations probe cooler outer disk 
regions and layers close to the midplane of circumstellar disks, observations in the 
mid-infrared (MIR) wavelength regime are more sensitive to warmer ($250\ \mathrm{K} \lse T \lse 1000\ \mathrm{K}$; 
see Schegerer et al.~2006: Fig.~1) disk regions, such as the surface of the inner regions where dust 
is directly irradiated by the central star. However, studies of the inner circumstellar
regions ($\rm \sim$$1\ \mathrm{AU}$) of objects in the closest star forming regions are only feasible by 
interferometric observations.  

Strong emission features in the MIR range at $\rm 10\ \mathrm{\mu m}$ and $\rm 20\ \mathrm{\mu m}$ 
corresponding to the Si-O stretching and bending modes of silicate grains, are assumed to result 
from absorption and reemission processes in optically thin dust layers of the
circumstellar disks. While silicate grains are expected to be initially amorphous and small 
($\lse 0.1\ \mathrm{\mu m}$; Mathis et al.~\cite{mrn}, MRN thereafter)\footnote{In this paper 
amorphous and small, i.e., primordial and interstellar dust grains, are called 
not-evolved/undeveloped.}, the crystallization of amorphous silicates starts at temperatures of 
$\sim$$1200\ \mathrm{K}$ (e.\/g., Gail~\cite{gail}). Moreover, high dust densities and turbulent 
processes in the interior of circumstellar disks favor dust grain growth to dust pebbles (e.\/g., 
Blum et al.~\cite{blum}; Johansen et al.~\cite{johansen}). The shape of the emitted 
silicate feature allows the estimation the predominant stage of the dust evolution in a young stellar 
object (YSO). Different degrees of crystallization and grain growth have already been shown in a 
large sample of T\,Tauri 
stars of different ages and stellar masses (e.\/g., Schegerer et 
al.~\cite{schegerer}). As temperature and density increase in circumstellar disks with decreasing 
distance to the central star, crystallinity and grain size sensitively depend on the radial position 
of the dust in a circumstellar system (e.\/g., Beckwith et al.~\cite{beckwithII}; 
Weidenschilling~\cite{weidenschilling}; Gail~\cite{gailII}: Fig.~28). In fact, observations with 
MID-infrared Interferometric instrument (MIDI) have already revealed a correlation between the radial position and the evolutionary stage of 
silicate dust in circumstellar disks around Herbig Ae/Be (HAeBe) stars, which are the more massive counterparts of 
T\,Tauri stars (Leinert et al.~\cite{leinertIV}; van Boekel et al.~\cite{boekelII}). 

The density, temperature and compositional structure of circumstellar dust disks and surrounding 
envelopes have been the central issue of many former studies (e.\/g., Chiang \& 
Goldreich~\cite{chiang}; D'Alessio et al.~\cite{dalessio}). Different modeling approaches have 
been tried to quantitatively explain and reproduce phenomenons like excess radiation, shapes of 
emission lines (e.\/g., Muzerolle et al.~\cite{muzerolle}; Natta et al.~\cite{nattaII}), flux 
variations (e.\/g., Herbig et al.~\cite{herbig}) and intensity distributions of images (e.\/g., Lucas et 
al.~\cite{lucas}). However, the evolution of (inner) disk structure and its correlation to dust 
evolution is still unclear (e.\/g., Millan-Gabet et al.~\cite{millan-gabetII}; Beckwith et 
al.~\cite{beckwithII}), and has been underestimated in actual modeling
approaches, or mainly reserved for theoretical studies
(e.\/g., Gail~\cite{gail}). However, interferometric observations in the
MIR wavelength range, which are now available, 
are sensitive to the inner disk structure where warm dust dominates. Including their sensitivity for 
the silicate feature, the correlation between inner disk structure and grain
evolution can be directly studied. 

In this paper we focus on modeling of the SED and spectrally resolved N band visibilities, which we 
obtained for the T\,Tauri star RY\,Tau with MIDI. The key questions of this study are the following: Is it 
possible to simultaneously model the SED and N band visibilities of RY\,Tau solely by an 
externally, i.e., passively heated, disk? Do different extensions of this model reproduce the 
observations, simultaneously? What do we learn about the (silicate) dust composition of the disk?

The result of previous measurements of RY\,Tau are presented in Sect.~\ref{section:previous measurements}. 
In Sect.~\ref{section:observation} we outline the observations of RY\,Tau and the subsequent data 
reduction. We present the radiative transfer code and the basic dust set of
our modeling approach in 
Sect.~\ref{section:modeling}. In the following Sect.~\ref{section:models} we compare the 
results of the different modeling approaches we used, i.e., the ative disk
model with and without an envelope, and point to supplements. In Sect.~\ref{section:dust 
composition} the dust composition of the upper disk layers and its dependence on the radial distance 
from the central star is studied. Finally, in Sect.~\ref{section:discussion} we draw comparisons 
between the used models, refer to previous results, including the models
of Akeson et al.~(\cite{akeson}) where near-infared (NIR) visibilites were modeled, and
discuss discrepancies. Furthermore, we investigate the possibility of the
existence of a stellar companion and compare RY\,Tau with HAeBe stars. 
Section~\ref{section:conclusion} summarizes our results. 

\section{Previous measurements}\label{section:previous measurements}
RY\,Tau, demonstrably observed for the first time in 1907
(Pickering~\cite{pickering}), is a well-known T\,Tauri 
star (Joy~\cite{joy}) that belongs to the Taurus-Auriga molecular cloud at a distance of 
$\sim$$140\ \mathrm{pc}$. Table~\ref{table:properties} shows the main properties of this object, 
which are obtained from previous measurements. Photometric fluxes are listed in 
Table~\ref{table:photometrie}. 
\begin{table}[!t]
  \caption{Observed properties of RY\,Tau. {\it References}~-- {\bf 1}: Perryman et 
    al.~(\cite{perryman}); {\bf 2}: Bertout et al.~(\cite{bertout}); {\bf 3}:
    Calvet et al.~(\cite{calvetII}); {\bf 4}: Mora et al.~(\cite{mora}); {\bf 5}: Beckwith et 
    al.~(\cite{beckwith}); {\bf 6}: Akeson et al.~(\cite{akeson}); {\bf 7}: Siess et 
    al.~(\cite{siess})}
  \label{table:properties}
  \centering
  \begin{tabular}{llr} \hline\hline
    Parameter & Value & Reference \\ \hline
    RA (J2000.0) & $04\ 21\ 57.4$ &  {\bf 1} \\
    DEC (J2000.0) & $+28\ 26\ 36$ & {\bf 1} \\
    Distance & $134^{+54}_{-31}\mathrm{pc}$ & {\bf 2} \\
    Visual Extinction & $(2.2 \pm 0.2)\ \mathrm{mag}$ & {\bf 3} \\ 
    Spectral type & F$8$\,III & {\bf 4} \\
    Stellar mass & $1.69\,\mathrm{M_{\odot}}$ & {\bf 5} \\
    Stellar luminosity & $12.8\ \mathrm{L_{\odot}}$ & {\bf 6} \\
    Accretion rate & $7.8 \times 10^{-8}\,\mathrm{M_{\odot}/yr}$ & {\bf 3}\\
    Age & $(6.5 \pm 0.9)\ \mathrm{Myr}$ & {\bf 7} \\ \hline
  \end{tabular}
\end{table}

RY\,Tau is a UX\,Ori-type star, i.e., this T\,Tauri object has revealed irregular photometric 
variability in the vi\-sible and NIR wavelength range. During several months in 
1983/84 and in 1996/97, its visible brightness increased from
$\sim$$11\mathrm{th}$ to $\sim$$9\mathrm{th}$ magnitude 
and decreased again to its initial value (Herbst \& Stine~\cite{herbst}; Zajtseva et 
al.~\cite{zajtseva}; Petrov et al.~\cite{petrov}; Herbst \& Shevchenko~\cite{herbstII}). Such a rare 
but strong variability is conventionally explained by variations of the obscuration of the 
central star caused by an inclined circumstellar disk and an envelope (e.\/g. Eiroa et 
al.~\cite{eiroa}). Smaller variations ($\Delta V \approx 0.1$, $\Delta J \approx 0.2$, $\Delta K 
\approx 0.2$) in the range of several days were also detected (Eiroa et al.~\cite{eiroa}). By a 
comparison between the maximum and minimum brightness of the object the photometric measurements, 
listed in Table~\ref{table:photometrie}, and our observations with MIDI
correspond to the ``quies\-cent'' state of the object, i.e., 
close to the photometric mini\-mum.

There is a wide range of values measured for the visual extinction $A_\mathrm{V}$ of RY\,Tau 
(Kuhi~\cite{kuhi}: $1.3\ \mathrm{mag}$; Cohen \& Kuhi~\cite{cohen}: $(1.9 \pm
0.2)\ \mathrm{mag}$; Strom et al.~\cite{strom}: $0.6\ \mathrm{mag}$; Beckwith
et al.~\cite{beckwith}: $2.7\ \mathrm{mag}$; Kenyon \& Hartmann~\cite{kenyon}:
$1.8\ \mathrm{mag}$). We adopt a 
value recently derived by Calvet et al.~(\cite{calvet}): $A_\mathrm{V}=(2.2
\pm 0.2)\ \mathrm{mag}$. The level of veiling in the visible range of the 
spectrum is low ($\lse 0.1$; Basri et al.~\cite{basri}; Hartigan et al.~\cite{hartiganII}; Petrov et 
al.~\cite{petrov}) but markedly higher in the infrared range ($>0,8$ ; Folha \& 
Emerson~\cite{folhaII}). 

A potential duplicity/multiplicity was not found by Leinert et al.~(\cite{leinertV}) by using NIR 
speckle interferometry reaching a spatial resolution between $0.13\arcsec$ and
$13\arcsec$ but the re\-gular variation of the photocenter, found by the 
astrometric measurements of HIPPARCOS, could be a  hint for a companion with a projected 
minimum distance of $3.27\ \mathrm{AU}$ ($23.6\,\mathrm{mas}$) and a position angle of $304^{\circ} 
\pm 34^{\circ}$ (Bertout et al.~\cite{bertout}). 
\begin{table}
  \caption{Photometric measurements of RY\,Tau. The data of different measurements are averaged and 
    the standard deviation is determined. For convience, all fluxes are given
    in Jansky and in magnitude. Conversion factors are taken from Leinert~(\cite{leinertI}). 
    {\it References}~-- {\bf 1}: Calvet et al.~(\cite{calvet}); {\bf 2}: Gezari-Catalog~(Gezari et
    al.~\cite{gezari}); {\bf 3}: 2\,MASS-Catalog~(Cutri et
    al.~\cite{cutri}); {\bf 4}: Rydgren et al.~(\cite{rydgren}); {\bf 5}: Elias~(\cite{elias}); 
    {\bf 6}: Hanner et al.~(\cite{hanner}); {\bf 7}: Strom et
    al.~(\cite{strom}); {\bf 8}: Harvey et al.~(\cite{harvey}); {\bf 9}: 
    Mora et al.~(\cite{mora}) }
  \label{table:photometrie}
  \centering
  \begin{tabular}{llr}\hline\hline
    wavelength [$\mathrm{\mu m}$] & flux [Jy] ([mag]) & Reference \\ \hline
    $0.36$ (U) & $0.04 \pm 0.01$ ($11.72 \pm 0.16$) & {\bf 1}\\
    $0.45$ (B) & $0.14 \pm 0.01$ ($11.23 \pm 0.08$) & {\bf 1}\\
    $0.55$ (V) & $0.33 \pm 0.01$ ($10.12 \pm 0.04$) &  {\bf 1}\\
    $0.64$ (R$_\mathrm{C}$) & $0.59 \pm 0.03$ ($9.3 \pm 0.05$) & {\bf 1}\\
    $0.79$ (I$_\mathrm{C}$) & $1.0 \pm 0.06$ ($8.52 \pm 0.06$) & {\bf 1}\\
    $1.25$ (J) & $1.6 \pm 0.8$ ($7.64 \pm 0.40$) & {\bf 2}, {\bf 3} \\
    $1.65$ (H) & $2.3 \pm 0.9$ ($6.48 \pm 0.30$) & {\bf 2}, {\bf 3} \\
    $2.20$ (K) & $3.8 \pm 0.5$ ($5.56 \pm 0.13$) & {\bf 2}, {\bf 3} \\ 
    $4.80$ (M) & $6.6 \pm 2.0$ ($3.4 \pm 0.28$) & {\bf 4}, {\bf 5} \\
    $11.0$ (N) & $20.0 \pm 0.3$ & {\bf 6} \\
    $25$ & $28 \pm 3$ & {\bf 7} \\
    $60$ & $18 \pm 4$ & {\bf 7} \\
    $100$ & $12 \pm 5$ & {\bf 7}, {\bf 8} \\
    $1300$ & $0.23 \pm 0.02$ & {\bf 9} \\ \hline
  \end{tabular}
\end{table}

\section{Interferometric observations and data reduction}\label{section:observation}
\subsection{Observing sequence}
RY\,Tau was observed with MIDI/VLTI (Very Large Telescope Interferometer; Leinert et al.~\cite{midi}) in 2004, November $1^\mathrm{st}$ 
and $4^\mathrm{th}$, within the scope of guaranteed time observations. The dates and universal 
times (UT) of the observations, as well as sky-projected baseline lengths (L) and position angles 
(PA) of the interferometer, are listed in Table~\ref{table:observation}. 
\begin{table}[b]
  \caption{\label{table:observation} Summary of the MIDI observations of
    RY\,Tau and calibrators. The dates, $UT$, $L$ (in m) and $PA$ (in degrees, 
    measured from North to East) of the sky-projected baselines are
    listed. The airmass $AM$ in the right column is given for the time of fringe
    tracking.  The observations
    with a projected baseline of $79\ 
    \mathrm{m}$ and $81\ \mathrm{m}$ provided an almost identical result (see 
    Fig.~\ref{figure:vis-data}).}
  \hfill{}\begin{tabular}{cccccc}
    \hline 
    \hline 
    Date & $UT$ & Object & $L$ & $PA$  & $AM$ \\
    \hline
    Nov. $1^\mathrm{st},\ 2004$ & 3:54 - 4:16 &  \object{HD\,25604} &
    $74$ &  $96$ & $1,68$ \\  
    Nov. $1^\mathrm{st},\ 2004$ & 4:37 - 4:56 &    RY\,Tau & $79$ &  $97$ &
    $1.83$ \\  
    Nov. $1^\mathrm{st},\ 2004$ & 4:58 - 5:07 &    RY\,Tau & $81$ &  $95$ &
    $1.79$ \\  
    Nov. $1^\mathrm{st},\ 2004$ & 5:54 - 6:10 &  \object{HD\,49161} & 
    $64$ &  $87$ & $1.56$\\   
    Nov. $1^\mathrm{st},\ 2004$ & 7:03 - 7:17 &  \object{HD\,31421} &
    $89$ &  $82$ & $1.28$\\  
    \\                 
    Nov. $4^\mathrm{th},\ 2004$ & 0:01 - 0:26 & \object{HD\,178345} &
    $57$ & $146$ & $1.42$\\
    Nov. $4^\mathrm{th},\ 2004$ & 2:19 - 2:47 & \object{HD\,188603} &
    $46$ & $169$ & $2.47$\\
    Nov. $4^\mathrm{th},\ 2004$ & 3:16 - 3:54 &  HD\,25604 & $61$ & $117$ &
    $1.75$ \\
    Nov. $4^\mathrm{th},\ 2004$ & 5:11 - 5:29 & \object{HD\,20644} &
    $59$ & $102$ & $1.69$ \\
    Nov. $4^\mathrm{th},\ 2004$ & 7:18 - 7:36 &  \object{HD\,37160} &
    $61$ & $107$ & $1.21$\\
    Nov. $4^\mathrm{th},\ 2004$ & 7:44 - 8:00 & RY\,Tau & $49$ &  $92$ &
    $1.95$\\
    Nov. $4^\mathrm{th},\ 2004$ & 9:00 - 9:23 &  \object{HD\,50778} &
    $61$ & $113$ & $1.04$\\
    \hline
  \end{tabular}\hfill{}
\end{table}
An observing sequence consists of the following steps:
\renewcommand{\labelenumi}{\roman{enumi}.}
\begin{enumerate}
\item Single-telescope imaging is used for a highly precise acquisition of the
  object within a field-of-view of $2\times2\arcsec$ in order to guarantee 
  a maximum overlap of the telescope beams. 
\item After the beam combiner is introduced into the optical path, the characteristic interference 
  pattern (fringe pattern) is found around an optical path difference (OPD) of zero. A low 
  resolution prism ($\rm \lambda / \delta \lambda \approx 30$), which is put in the optical path of 
  the combined beams, allows us to obtain spectrally resolved interferograms and the 
  wavelength-dependent correlated flux $F_\mathrm{corr} ( \lambda )$.
\item The spectrum $F_\mathrm{total}( \lambda )$ is determined by single-telescope exposures that 
  are recorded on the same detector pixels as the fringe signal $F_\mathrm{corr} ( \lambda )$. By 
  definition, the spectrally resolved visibility $V(\lambda)$ is obtained as the ratio of the 
  correlated and the total flux 
  \begin{eqnarray}
    \hfill{}
    V(\lambda)=\frac{F_\mathrm{corr}( \lambda )}{F_\mathrm{total}(\lambda)}.
    \hfill{}
    \label{eq: visibility}
  \end{eqnarray}
  This visibility is still biased by instrumental effects.
\item The transfer function of the instrument is determined by the observation of calibrator stars 
  before and after the observation of the scientific target. A known transfer
  function is required for the elimination of the instrumental and atmospheric influences. The 
  calibrators are selected for their known diameter, the absence of strong photometric variabilities 
  and companions, and their angular distance to the scientific target. Some of the 
  calibrators were also used for absolute flux calibration. The error of our calibrated visibility 
  $V(\lambda)$ is the $1\ \sigma$ deviation that is obtained by the observations of different 
  calibrators in one night (see Table~\ref{table:observation}). 
\end{enumerate} 
Instrumental informations and observing procedures are described by Leinert et 
al.~(\cite{leinertII}; \cite{leinertIII}; \cite{leinertIV}) and Ratzka~(\cite{ratzka}). 

\subsection{Data reduction}
The reduction procedure of MIDI data is complex and has been described in detail by Leinert et 
al.~(\cite{leinertIV}), Ratzka~(\cite{ratzka}), and Jaffe~(\cite{jaffe}). The
data obtained with MIDI were 
reduced with the MIA software that is based on power spectrum analysis and the results were 
independently confirmed by using the EWS software. The EWS software contains a coherent integration 
algorithm, which involves a kind of shift-and-add in the complex plane. Both reduction software 
packages are publicly available.\footnote{{\tt http://www.mpia-hd.mpg.de/MIDISOFT/} and \\
  {\tt http://www.strw.leidenuniv.nl/$^\sim$koehler/MIA+EWS-Manual/}}
 
\subsection{Observational results}
\begin{figure}[!htp]
\resizebox{\hsize}{!}{\includegraphics{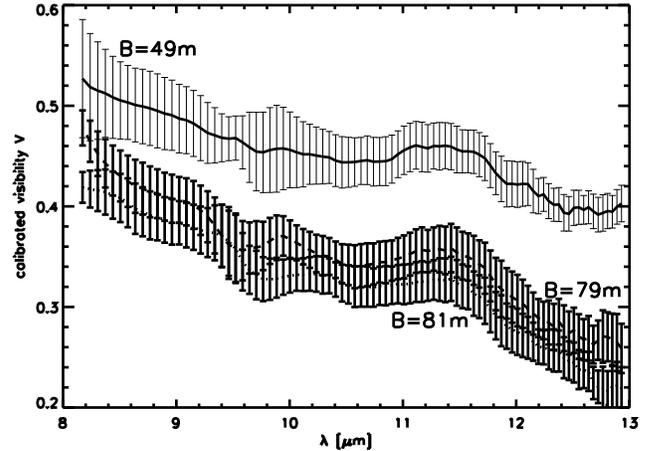}}
\caption{The spectrally resolved, calibrated visibility data derived from our MIDI observations 
  (Table~\ref{table:observation}). The error bars are the $1\ \sigma$ deviations that result from 
  the observations of different calibrators in one night.}
\label{figure:vis-data}
\end{figure}

The resulting wavelength dependent visibility curves for the three baselines are shown in 
Fig.~\ref{figure:vis-data}, including $1\ \sigma$ error bars. The spectrophotometry of the silicate 
emission band $F_\mathrm{total}(\lambda)$, obtained during the measurements, 
is shown in Fig.~\ref{figure:fitting}, and also included in the SED of RY\,Tau 
(Fig.~\ref{figure:ry}). Figure~\ref{figure:fitting} also shows the observed correlated fluxes 
$F_\mathrm{corr}( \lambda )$, formally obtained as a product of these two quantities 
(Eq.~\ref{eq: visibility}). We refer to the near-coincidence of the observations at baseline lengths 
of $79\ \mathrm{m}$ and $81\ \mathrm{m}$. 

\section{Tools}\label{section:modeling}

\subsection{MC3D -- Monte Carlo code for radiative transfer}\label{section:mc3d}
In contrast to many previous investigations where the radiative transfer function has been 
solved (e. g., Sonnhalter et al.~\cite{sonnhalter}; Chiang \& Goldreich~\cite{chiang}; 
Dullemond et al.~\cite{dullemond}), we use the well-tested code 
MC3D, which is based on the Monte-Carlo method (Wolf et al.~\cite{wolfI}; Pascucci et 
al.~\cite{pascucci}). Considering an axially symmetric object, we assume a two-dimensional geometry 
in a polar coordinate system ($r$,$\theta$) with a logarithmic grid spa\-cing in $r$ and a 
uniform grid spacing in $\theta$. Heating sources like the central star, accretion effects, and 
heated dust grains determine the temperature distribution. The flux of the
central star is determined by the theoretical stellar atmosphere model provided by the
Kurucz~(\cite{kurucz}). The product of the dust-specific 
absorption efficiency $Q_\mathrm{abs}(\lambda)$, the 
grain surface and the blackbody emission $B_{\nu}(T_\mathrm{dust})$ represent the flux that a dust 
grain with a temperature of $T_\mathrm{dust}$ reemits in local thermal equilibrium (Kirchhoff's law 
of local thermal equilibrium).  Sources like the central star and accretion are treated as blackbody 
emitters. Gas molecules and atoms are not considered in our models. 

The radial density distribution of the disk is given by the surface density profile 
\begin{eqnarray}
\hfill{}
\Sigma(r)=\Sigma_\mathrm{0} \cdot r^{-p}
\hfill{}
\label{eq: sigma}
\end{eqnarray}
with the radial coordinate $r$, an exponent $p$ and a constant $\Sigma_\mathrm{0} = 
100\ \mathrm{g\,cm^{-2}}$ (Weidenschilling~\cite{weidenschillingII}). The vertical 
density distribution is calculated self-consistently assuming hydrostatic equilibrium, i.e., the 
balance of gravitational and thermal pressure (Schegerer et al., in prep.). 

After temperature and density distribution have been determined, the SED and the projected image of 
the star and its circumstellar environment, considering an inclination angle $\vartheta$,  are 
calculated. The resolution of the image is by a factor of $\sim$$10$ better than the 
resolution of the observations. For projected baselines of $79\ \mathrm{m}$ and $49\ \mathrm{m}$, our 
observations with MIDI reached spatial resolutions of $\sim$$1.8\ \mathrm{AU}$ and $\sim$$2.8\ \mathrm{AU}$ at a 
distance of $\sim$$140\ \mathrm{pc}$.\footnote{The spatial resolution power $R$
  of an interferometer is given by the ratio of the observing wavelength
  $\lambda$ and the sky-projected distance between a telescope pair, i.e., the
  effective baseline length $B$: $R=\lambda / (2B)$.}

\subsection{Dust model}\label{section:dust}
The infrared excess that is emitted from YSOs originates from heated dust in the circumstellar 
environment.  Assuming compact, homogeneous, and spherical dust grains, their
optical properties, such as 
scattering and extinction cross sections, are determined by Mie scattering-theory from the measured 
complex refractive index of the specific material (Bohren \& Huffman~\cite{bohren}). In our 
modeling approach, we assume a dust mixture of ``astronomical silicate'' and graphite with the 
relative abundances of $62.5$\% for astronomical silicate and $37.5$\% for graphite (Draine \& 
Malhotra~\cite{draineII}). The dielectric function of astronomical silicate was formerly synthesized 
by Draine \& Lee~(\cite{draine}) in order to reproduce the extinction of
different silicate compounds in interstellar 
space. We consider an improved version of this dielectric function (Weingartner \& 
Draine~\cite{weingartner})\footnote
{
See {\tt http://www.astro.princeton.edu/$\rm \sim$draine}.
}, 
which was recently confirmed by a study of the interstellar extinction in the NIR wavelength 
range (Indebetouw et al.~\cite{indebetouw}). For graphite we adopt the 
$\frac{1}{3}:\frac{2}{3}$ ratio with $\kappa_{\nu}=[\kappa_{\nu}(\epsilon_{\|}) + 2 
\kappa_{\nu}(\epsilon_{\bot})]/3$, where $\epsilon_{\|}$ and $\epsilon_{\bot}$ are the components 
of the graphite dielectric tensor for the electric field parallel and perpendicular to the 
crystallographic c-axis and $\kappa_{\nu}$ is the mass absorption coefficient. The strongly 
absorbing graphite grains efficiently contribute to the heating of the dusty circumstellar 
environment. The ratio of the extinction efficiency factor of carbon dust to
the extinction 
efficiency 
factor of silicate dust is $\sim$$10$ in the NIR wavelength range (Draine \&
Lee~\cite{draine}; Wolf \& 
Hillenbrand~\cite{wolfII}). 

We consider a grain size power law $n(a) \propto a^{-3.5}$ with $a_\mathrm{min} \leqq a \leqq 
a_\mathrm{max}$, where $n(a)$ is the number of dust particles with radius $a$. This power law was 
formerly found by MRN studying extinction of interstellar carbon and silicate with ty\-pical sizes 
between $0.005 \mathrm{\mu m} - 0.01 \mathrm{\mu m}$ and $0.025\ \mathrm{\mu m} - 0.25 \mathrm{\mu 
m}$, respectively. This grain size power law has already been used in former modeling 
aproaches of YSOs. We use a minimum particle size of $a_\mathrm{min} = 0.005\ 
\mathrm{\mu m}$ in all of our modeling approaches. 

The maximum grain size $a_\mathrm{max}$ affects the mass absorption coefficient 
$\kappa_\mathrm{\nu}(a)$ of dust, i.e., $\kappa_\mathrm{\nu}(a)$ increases with $a_\mathrm{max}$ for 
sizes up to a few 
mm. The mm flux depends on the disk mass $M_\mathrm{disk}$ and the absorption coefficient 
$\kappa_\mathrm{\nu}(a)$, i.e., $a_\mathrm{max}$, in parti\-cular, when assuming an 
optically thin disk in the mm wavelength range. Correspondingly, the spectral index $\alpha$ 
($F_\mathrm{\nu} \propto \nu^{\alpha}$) decreases with an increase of $a_\mathrm{max}$ from an 
absolute value of $\sim$$4$ (for $a_\mathrm{max} < 0.1\ \mathrm{\mu
  \mathrm{m}}$) to $\sim$$2$ (only for 
large bodies). The correlation between the spectral index $\alpha$ and 
the maximum dust grain size $a_\mathrm{max}$ was formerly studied by D'Alessio et 
al.~(\cite{dalessio}), while Wood et al.~(\cite{woodII}) also investigated the correlation between disk 
mass and mm flux. 

In our modeling approach, we have found that a dust distribution with a maximum grain size of 
$a_\mathrm{max}=0.25 \mathrm{\mu m}$ and the above mentioned grains size power
law for silicate and carbon generally underestimates the mm flux unless a disk mass in the 
range of $\sim$$1.0\ \mathrm{M_{\odot}}$ is assumed. Circumstellar disks with such high masses 
are potentially gravitationally unstable (e.\/g., Laughlin \&
Bodenheimer~\cite{laughlin}; Boss~\cite{boss}; Lodato \& 
Bertin~\cite{lodato}). Moreover, an upper grain size of $a_\mathrm{max}=0.25\ \mathrm{\mu m}$ results 
in a too steep mm slope, in contrast to the measured spectral index $\alpha$. Using the Very Large 
Array for their mm measurements, Rodmann et al.~(\cite{rodmann}) found a spectral index of $\alpha = 
2.55 \pm 0.09$ for RY\,Tau and derived a maximum grain size of $a_\mathrm{max}=1\ \mathrm{mm}$. 
The latter maximum grain size is used in our modeling approach. We have to mention that the spectral 
index $\alpha$ provides only a lower limit for the maximum dust size as it converges for 
$a_\mathrm{max} > 1\ \mathrm{mm}$. 

Although the mm wavelength range of the observed SED can sufficiently be simulated considering grain 
sizes up to $1\ \mathrm{mm}$ and relatively low disk masses ($< 10^{-2}\ \mathrm{M_{\odot}}$), the 
spectral contribution in the NIR wavelength range strongly decreases with increasing maximum grain 
size. The dust particles with $a_\mathrm{max} > 1\ \mathrm{\mu m}$ can be less effectively heated 
than the smaller particles of the canonical MRN distribution.\footnote{For all different dust sets 
we assume a constant exponent of $-3.5$ for the grain size power law $n(a)$.}
This effect is the reason why we implement a 
two-layer dust model in our mo\-de\-ling approach. The disk interior contains
a maximum dust grain size 
of $1\ \mathrm{mm}$ while the MRN grain size distribution with $a_\mathrm{max} = 0.025\ \mathrm{\mu 
m}$ is used in the upper disk layers where the optical depth $\tau_\mathrm{N}$
in N band, measured vertical to the disk midplane, falls below unity. Such a division of the disk 
is based on the idea of the favored settling of larger dust grains. Furthermore, dust particles are 
assumed to mainly grow in the denser regions of the disk close to the midplane
(e.\/g., Lissauer~\cite{lissauer}; Blum \& Wurm~\cite{blum}). Similar disk models with two or more different 
dust layers have already been proposed by Chiang \& Goldreich~(\cite{chiang}) and used by Whitney et 
al.~(\cite{whitney}), for instance. However, it is still an open question how strongly dust grains are mixed in 
the circumstellar environment (e.\/g., D'Alessio et al.~\cite{dalessioII}; Gail~\cite{gail}; McCabe et 
al.~\cite{cabe}; Wolf et al.~\cite{wolfIV}).

In order not to determine the temperature distribution of each single dust component and to 
accelerate the radiative transfer simulations, we construct an ``artificial'' particle 
with optical constants that are derived by averaging the optical pro\-perties of carbon and 
astronomical silicate of different sizes in each dust layer. Such an approach was justified by Wolf~(\cite{wolfIII}). 

\section{Models of the density structure}\label{section:models}
In the following subsections we will present our approaches to model the SED and the MIR 
visibilities that we obtained with MIDI. An active disk model is our favorite  
approach. Such a model was also used to reproduce SED and K Band visibilities of RY\,Tau obtained 
with the Palomar Testbed Interferometer (Akeson et al.~\cite{akeson}). Additionally, 
an active disk model with an envelope and potential supplements are discussed. One
of the main issues of this study is to clarify if the 
models we use can be distinguished.

The existence of a passively-heated circumstellar disk in our analysis is beyond all questions. The 
paradigm that the formation of a disk is one of the evolutionary stages of circumstellar stuctures 
has been finally confirmed by the images of many YSOs in different wavelength ranges (e.\/g., Padgett 
et al.~\cite{padgett}; Mannings \& Sargent~\cite{mannings},~\cite{manningsII}). The T\,Tauri star 
RY\,Tau is a Class II object (Kenyon \& Hartmann~\cite{kenyon}) where a 
surrounding disk has already formed. Therefore, a passively heated disk is a basic ingredient 
and will be retained in our different modeling approaches.

The disk model is characterized by the disk mass $M_\mathrm{disk}$\footnote
{
We assume a gas-to-dust mass ratio of 100:1.
}, the inner $R_\mathrm{in}$ and outer disk radius $R_\mathrm{out}$, the inclination angle 
$\vartheta$, and an exponent $p$, which represents the radial dependence of the surface density 
$\Sigma$ (see Eq.~\ref{eq: sigma}). The inner radius $R_\mathrm{in}$ is given in advance, but can 
also be considered as a starting value that is iterated until the temperature at the inner radius 
falls below the the sublimation temperature of $1600\ \mathrm{K}$ (Duschl et al.~\cite{duschl}). 
Properties of the central star, like the stellar temperature $T_\mathrm{\star}$, stellar luminosity 
$L_\mathrm{\star}$ and stellar mass $M_\mathrm{\star}$, are additional model parameters but these 
quantities are well constrained by previous studies (see Table~\ref{table:properties}). 

\subsection{Active disk}\label{section: accretion} 
Our active disk model is a passively heated (dust) disk where accretion effects are 
added. The existing MC3D radiative transfer code therefore has to be extended. 
The implemented accretion effects are briefly described in
Appendix~\ref{appendix}.

Apart from the parameters of the disk and the star, our accretion model requires three additional 
model parameters: accretion rate $\dot{M}$, boundary temperature $T_\mathrm{bnd}$ of the accreting 
regions on the surface of the star, and a magnetic truncation radius $R_\mathrm{bnd}$. Defining 
the inner radius of the gaseous disk inside $R_\mathrm{in}$, K\"onigl~(\cite{koenigl}) showed that 
the truncation radius is not an independent quantity, but depends on stellar radius, mass, accretion 
rate and magnetic field strength. However, as we do not know the exact magnetic field strength of 
RY\,Tau, we fix the boundary temperature and truncation radius to $8000\ \mathrm{K}$ and 
$5\ R_{\star}$, respectively. Both values were already used for the same object by the study of 
Akeson et al.~(\cite{akeson}) and were justified by the assumed large magnetic field of RY\,Tau in 
the range of a few kilo-Gauss. 

The best fit parameters for this model are given in Table~\ref{table:ry} and the model is compared 
to the observations in Fig.~\ref{figure:ry}. The accretion luminosity is $1.2\ \mathrm{L_\mathrm{\odot}}$. 
This model suffers from the following deficiency: the far-infrared (FIR) wavelength range in the SED is 
slightly overestimated. A potential improvement of this active disk model could be a 
``truncated outer disk'', formerly suggested by Lucas \& Roche~(\cite{lucas}) and recently used by 
Preibisch et al.~(\cite{preibisch}). For this, the primary density distribution of the disk is 
truncated at an outer radius $R_\mathrm{TD}$ by multiplying the surface density $\Sigma(r)$ 
(Eq.~\ref{eq: sigma}) with a Fermi-type function. With a constant $C_\mathrm{TD}>0$
\begin{eqnarray}
  \Sigma_\mathrm{TD}(r) = \frac{\Sigma(r)}{1+\exp(C_\mathrm{TD}\frac{r}{R_\mathrm{TD}} - C_\mathrm{TD})}
\label{eq:truncation}
\end{eqnarray}
This decreases the extended IR emission while the inner disk structure is modified much 
less than the outer disk regions. It could have its physical origin in a stellar companion which
truncates the outer disk region. In fact, Bertout et al.~(\cite{bertout}) pointed to a potential 
binarity of the system. Our study has shown that only a truncation radius of
the order of $10\ \mathrm{AU}$ could effectively decrease the computed
SED (Preibisch et al.~\cite{preibisch}). However, this truncation radius is still too large to correspond to
the findings of Bertout et al.~(\cite{bertout}). Moreover, such a truncation radius is so small that
the MIR visibilities increase. 
Another possibility to decrease the extended IR emission is dust settling that results in a 
flattening of the flared disk as suggested by Miyake \& Nakagawa~(\cite{miyake}) and Dullemond \& 
Dominik~(\cite{dullemondIII}). As outer disk regions are less heated, a less effective flared disk 
would also decrease the visibility. 

Studying SED and visibility, the inclination $\vartheta$ and the position angle $PA$ of the object 
cannot be clearly derived. Two visibility points are not sufficient to derive these values. With 
respect to the SED only an upper limit of $\lse$$65^{\circ}$ can be determined. This corresponds to 
the angle of the line of sight where the optical depth $\tau_\mathrm{V}$ in the model exceeds unity. 

\begin{figure}[!htp]
\resizebox{\hsize}{!}{\includegraphics[scale=0.52]{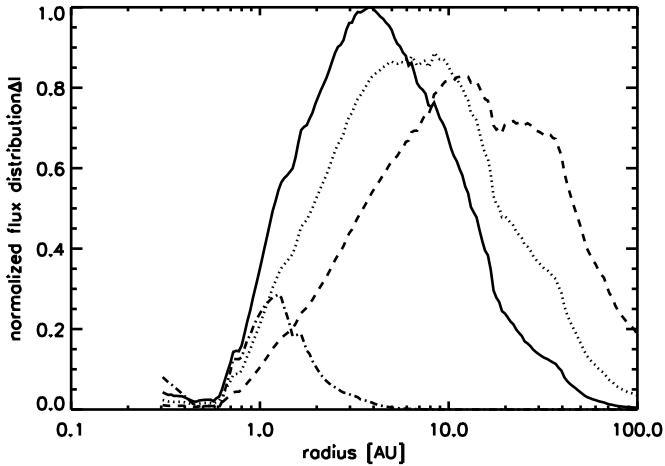}}
\caption{Radial flux distribution $ \Delta I / \Delta r $ for a wavelength of $8\ \mathrm{\mu m}$ 
  (solid), $10\ \mathrm{\mu m}$ (dotted), $12\ \mathrm{\mu m}$ (dashed curve)
  and $2.4\ \mathrm{\mu m}$ (dashed-dotted curve). The mean flux 
  $I$, emitted at a radius $r$, is multiplied with $4 \pi r^2$ and normalized by the 
  maximum of the radial flux distribution $I$ at a wavelength of $\lambda = 8\ \mathrm{\mu m}$.}
\label{figure:leinert}
\end{figure}

The outer disk radius $R_\mathrm{out}=270\ \mathrm{AU}$ of this model is
larger than the results from previous measurements. Andrews \&
Williams~(\cite{andrews}) derived an outer 
disk radius of $150\ \mathrm{AU}$ with the Submillimeter Array (SMA), while Rodmann et 
al.~(\cite{rodmann}) found an outer disk radius of only $(90 \pm 15)\ \mathrm{AU}$ with 7-mm continuum 
observations at the Very Large Array (VLA). Fig.~\ref{figure:leinert} shows at
which stellar distances the MIR flux arises in our model. This 
radial flux distribution confirms that it has its origin in the inner disk regions ($< 10\ 
\mathrm{AU}$), mainly, that can only be observed with interferometric methods. But it also shows that 
MIDI is not sensitive to outer disk regions ($> 40\ \mathrm{AU}$), including
the outer disk radius. A modeling of mm maps of this object allows to
determine the outer disk radius but this is out of the scope of this
paper. Additionally, it should be underlined that the distribution of the NIR flux
from the disk culminates at $\sim$$1\ \mathrm{AU}$ for this model.   

\begin{figure*}
\resizebox{0.48\hsize}{!}{\includegraphics{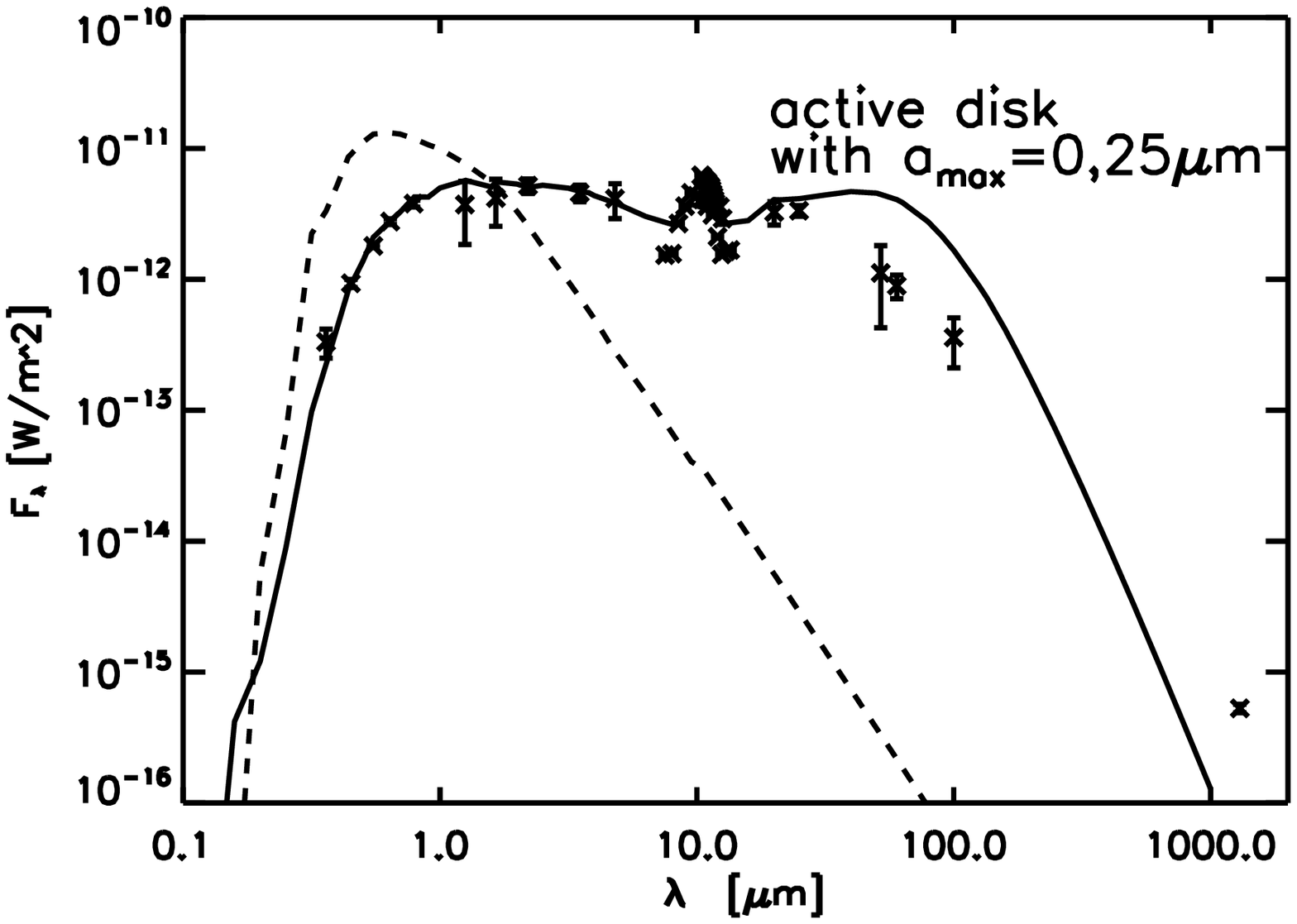}}
\resizebox{0.48\hsize}{!}{\includegraphics{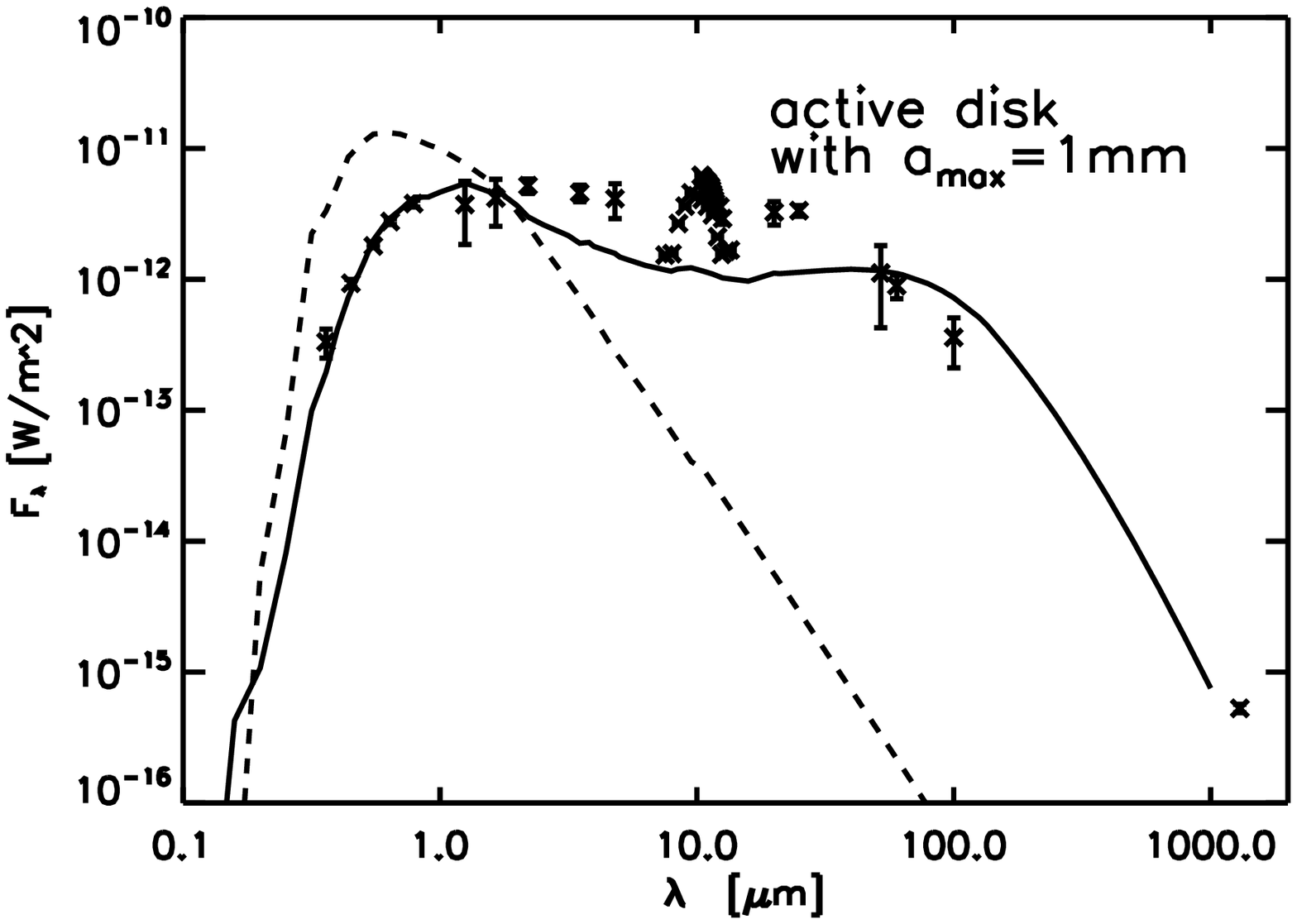}}
\caption{{\it Left:} Deficiency of the active disk model at long
  wavelength if only dust with standard size MRN distribution is assumed
  ($a_\mathrm{max}=0.25\ \mathrm{\mu m}$; compare middle panel in
  Fig.~\ref{figure:ry}). Observations are represented by crosses with error
  bars (s. Table~\ref{table:photometrie}). The predicted slope into 
  the mm range is too steep. The dashed line shows the unreddened stellar
  flux. {\it Right:} The same as the left figure but an maximum dust grain
  radius of $a_\mathrm{max}=0.25\ \mathrm{\mu m}$ is assumed.}
\label{figure:onedustlayer}
\end{figure*}

The disk mass of this model $M_\mathrm{disk} = 4 \times 10^{-3}\
\mathrm{M_{\odot}}$ is by a factor of $3$ smaller than the value that was
found by Akeson et al.~(\cite{akeson}). In this context we have to underline that 
the disk mass strongly depends on the dust set that is used in the model (see 
Sect.~\ref{section:dust}). The Fig.~\ref{figure:onedustlayer}, left, shows the SED for our 
active disk model but with the MRN dust set ($a_\mathrm{max}=0.25\
\mathrm{\mu m}$), only. All other model parameters are adapted. In this figure
it can be seen that the resulting flux in the mm wavelength range strongly declines for 
$a_\mathrm{max}=0.25\ \mathrm{\mu m}$.  

Finally, we attempted to reproduce SED and N band visibilities with a passive 
disk model, solely, instead of implementing accretion effects, additionally. Such passive disk 
models for different parameter sets were calculated but without reproducing the SED and MIR 
visibilities, sufficiently. Each modeled SED of these simulations suffers from similar deficiencies 
(see upper panel in Fig.~\ref{figure:ry}): in contrast to the photometric measurements the predicted 
NIR flux between $\lambda \approx 3\ \mathrm{\mu m}$ and $8\ \mathrm{\mu m}$ is generally 
underestimated. Moreover, the model is too strongly spatially resolved (visibilities which are 
too low) in
comparison with the measured visibilities. 

This lack of NIR radiation in
``naked'' passive disk models was reported by Hartmann et al. in \cite{hartmann}. 
Obviously, accretion can generate the missing NIR excess in a region which cannot resolved by 
the interferometer. In the following subsections we 
present two further modifications (the puffed-up inner rim wall, an
envelope) which have been considered to reproduce distinct NIR excess. 

\subsection{Inner rim wall}\label{section: inner rim}
In addition to an accretion model that is described in Sect.~\ref{section: accretion}, Natta 
et al.~(\cite{natta}) suggested that a puffed-up wall at the inner disk rim 
could also produce an increased NIR excess. Dust at the inner edge of the disk is strongly heated by 
direct stellar irradiation. Based on the idea of hydrostatical equilibrium this heating of the inner 
disk edge is assumed to cause an expansion of the dust layers perpendicular to the midplane of the 
disk. Dullemond et al.~(\cite{dullemond}) established an analytical model of the 
puffed-up, inner rim wall as a supplement to the Chiang-Goldreich model (Chiang \& 
Goldreich~\cite{chiang}) and defined the ``inner rim scale-height'' $H_\mathrm{rim}$ as follows: 
\begin{eqnarray}
\hfill{}
H_\mathrm{rim}=\chi_\mathrm{rim}h_\mathrm{in},
\hfill{} 
\label{rim}
\end{eqnarray}
where $\chi_\mathrm{rim}$ is a dimensionless factor, larger than unity. The quantity 
$h_\mathrm{in}$ represents the scale-height of the inner disk without extra vertical 
expansion.\footnote
{
The scale-height is defined as the vertical distance from the midplane where the density has 
decreased by a factor $e \approx 2.718$ (Euler's constant).
}
Simulations of young stellar objects (YSOs) where the inner rim wall has been successfully applied, have been 
published by Dominik et al.~(\cite{dominik}), Pontoppidan et al.~(\cite{pontoppidan}), and Eisner et 
al.~(\cite{eisnerII}), for instance. The reproduction of interferometric observations in the NIR 
regime assuming simple, analytical ring models (Millan-Gabet et al.~\cite{millan-gabet}) has also 
been used as a further confirmation of the inner rim wall (Natta et al.~\cite{natta}). 
Furthermore, a puffed-up inner rim, shadows adjacent areas 
of the disk from direct stellar radiation (``self-shadowed disk''). This effect is used to explain 
the spectral shape of the FIR excess from YSOs and to explain the classification of Group I 
and Group II sources (Meeus et al.~\cite{meeus}; Dullemond \& Dominik~\cite{dullemondII}). 
Arising from a flared disk a Group I source has a flat SED over the entire infrared wavelength 
range. The SED of a Group II source, however, strongly declines towards the
FIR. Such a decline 
was explained by a self-shadowed disk.

The puffed-up inner rim wall is still a controversial topic (e.\/g., Millan-Gabet et 
al.~\cite{millan-gabetII}). In particular, it was shown that a puffed-up inner rim wall does not 
generally emit enough radiation to cause the observed NIR excess in contrast to an  
{\it envelope} similar to the one that we implement in Sect.~\ref{section: dusty envelope} 
(Vinkovi\'c et al.~\cite{vinkovic}). Another open issue is the static stability of the 
proposed sharp, inner rim wall (Dullemond et
al.~\cite{dullemond}). Considering the sublimation temperature of the used
dust species, Isella \& Natta (\cite{isella}) 
revised the previous inner rim model by a more rounded-off inner rim.  Monnier et al.~\cite{monnier} 
observed RY\,Tau in the NIR wavelength range using the Infrared Optical Telescope Array (IOTA) where 
a spatial resolution of $\gse$$1\ \mathrm{AU}$ has been reached, comparable to our MIDI observations. 
In fact, their modeling results were incompatible with the models possessing vertical inner walls. 

\begin{figure}[!htp]
\resizebox{\hsize}{!}{\includegraphics[scale=0.31]{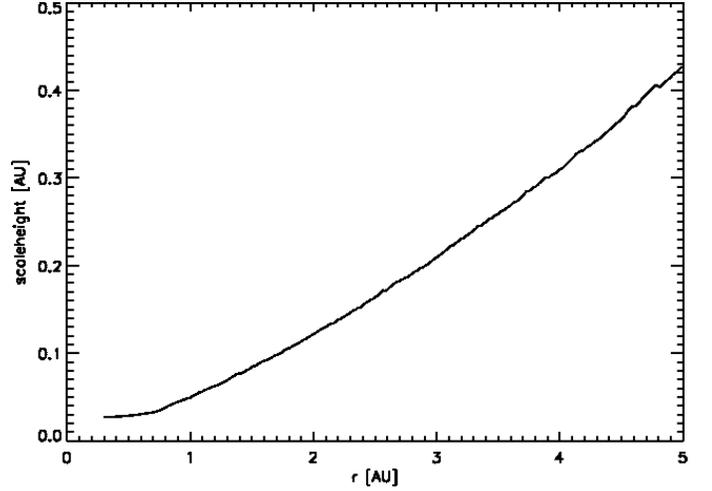}}
\caption{Scaleheight versus disk radius of our disk model from Sect.~\ref{section: accretion}. We do not see a 
  puffed-up inner rim in the sense of a local maximum of the scaleheight at the inner rim 
  and a following local minimum at slightly larger radii although the rim of
  our model catches a large fraction of stellar radiation. }
\label{figure:scaleheight}
\end{figure}

In our model the vertical density distribution is calculated assuming hydrostatic equilibrium. 
Therefore, the potential formation of a puffed-up inner wall is included in a natural way. However, 
it requires special computational care. In 
order to detect the effect of the puffed-up inner rim wall the size of grid cells in the inner 
region of the disk model should be small enough. A too coarsely meshed grid results in too low, 
averaged cell temperatures and in the absence of a potential puffed-up inner rim wall. We use a 
polar coordinate system ($r$, $\theta$) in our two-dimensional model with uniform steps in $\theta$ 
($\Delta \theta = 1^{\circ}$) and a logarithmic scale for $r$ (most inner step $\Delta r \approx 
0.01\ \mathrm{AU}$). Therefore, the inner grid cells have an approximate size of $\sim$$(0.01 \times
0.005)\ \mathrm{AU^2}$. According to Dullemond et
al.~(\cite{dullemond}) a puffed-up scaleheight of 
$0.05\ \mathrm{AU}$ up to $0.10\ \mathrm{AU}$ can be expected. Figure~\ref{figure:scaleheight} 
shows the scale-height of our active disk model of Sect.~\ref{section: accretion}. 

The fact that we do not see an excessively puffed-up inner wall in our approach does not 
necessarily exclude such a phenomenon. It is possible that the T\,Tauri star RY\,Tau is 
still too faint. In their study Dullemond \& Dominik~(\cite{dullemondII}) assumed HAeBe stars with 
temperatures in the range of $T_\mathrm{eff}=10,000\ \mathrm{K}$ and luminosities up to several $10$ 
solar luminosities (see also Dominik et 
al.~\cite{dominik}). The optical depth of the inner disk region also affects
the formation of a rim wall. If the optical depth is small at the inner disk
region, the radiation is not absorbed mainly at the inner rim of the disk but
on a larger scale. In this context Dullemond \& Dominik~(\cite{dullemondII}) showed
that the vertical height of the inner rim wall is notably boosted for
exponents $p \approx 4$ for the surface density (Eq.~\ref{eq: sigma}). Finally,
the properties of the dust grains that are used in the 
modeling approach can also affect the inner rim as Vinkovi\'c et
al.~(\cite{vinkovic}) mentioned. In Fig.~\ref{figure:leinert} it is shown that
the inner rim of our model emits only a smaller fraction of the NIR radiation.    

Our simulations are based on a {\it Monte Carlo} approach. Its drawback is the presence 
of statistical noise in the result which could in principle blur an effect such as the puffed-up 
inner rim wall. In order to 
quantify an upper limit of the scaleheight of a potential puffed-up inner rim,
we assume 
that the 
scaleheight at the innermost $0.3-3\ \mathrm{AU}$ can be described by a quadratic polynomial ($h 
\propto r^2$). This polynomial is then fitted to the scaleheight. The corresponding standard 
deviation $\sigma_\mathrm{SD}$ between the quadratic function and the scaleheight represents the 
maximum height of a potential puffed-up inner rim. We get $\sigma_\mathrm{SD} \lse 0.002\ 
\mathrm{AU}$ which is much smaller than the value derived by Dullemond et al.~(\cite{dullemond}).
 
The strength of the inner rim wall is still an open issue. It depends strongly on dust 
properties, radial disk structure and stellar properties that are used in the models. Future, 
highly spatially resolved observations in the NIR wavelength regime, that are
sensitive to the hot inner 
edge of the disk, will us allow to decide to what extent the effect of a puffed-up inner disk wall 
exists and which observational effects are provoked by this phenomenon. 

\subsection{Active disk with a dusty envelope}\label{section: dusty envelope}
Many studies (e.\/g., Hartmann et al.~\cite{hartmann}; Calvet et al.~\cite{calvet}) have shown 
that a dusty envelope around YSOs and Class-I sources, in particular, substantially contributes to 
the observed NIR excess. With respect to its optical depth a dusty envelope could even dim the 
stellar radiation in the visible wavelength regime. 

There are also several studies which justify an envelope structure around the star and disk of 
RY\,Tau. Vink et al.~(\cite{vink}) found that changes of the polarization across the $\mathrm{H 
\alpha}$ line of RY\,Tau are based on scattering effects due to an extended dusty envelope. Direct 
evidence for such a circumstellar halo has been provided by R and I coronagraphic, large-scale images 
of RY\,Tau (Nakajima \& Golimowski~\cite{nakajima}), in particular, and NIR, scattered light images 
around different YSOs (e.\/g., Padgett et al.~\cite{padgett}; Allen et al.~\cite{allen}), in general. 
Certainly, one challenge for interferometric studies is to decide whether the observed large-scale 
halo around RY\,Tau extends down to the inner disk region. 

In the context of axisymmetric accretion models Ulrich~(\cite{ulrich}) created an infall model of 
circumstellar gas and dust in an envelope structure in order to reproduce the emission-line 
$\mathrm{H}\alpha$ and $\mathrm{H}\beta$ profiles of type I/II P Cygni objects. This ansatz has been 
successfully used in modeling infrared images of Class-I objects (e.\/g., Lucas \& Roche~\cite{lucas}; 
Wolf et al.~\cite{wolfIV}). In contrast to Ulrich's approach we add a more simple spherical dust 
confi\-guration to the disk model. The spherical envelope in our model is geometrically constrained by 
the inner ($R_\mathrm{in}$) and outer ($R_\mathrm{out}$) disk radius. With the density distributions 
of the disk $\rho_\mathrm{disk}(r,\theta)$, of the envelope $\rho_\mathrm{env}(r,\theta)$, and the 
the position vector $\vec{r}$ as well as the coordinates $r$ and $\theta$ we define 
\begin{eqnarray}
\hfill{}
\rho_\mathrm{env}(r,0) = c_{1} \cdot \rho_\mathrm{disk}(R_\mathrm{in},0) 
\cdot \left(\frac{|\vec{r}|}{R_\mathrm{in}}\right)^{-c_{2}}
\hfill{}
\label{eq: envelope}
\end{eqnarray} 
where $c_\mathrm{1}<<1$ and $c_\mathrm{2} > 0$. The constraint $c_\mathrm{1}<<1$ guarantees a low 
optical depth of the envelope and the possibility to observe the innermost region of the disk. Disk 
and envelope are combined by 
\begin{eqnarray}
\hfill{}
\rho(\vec{r}) = & \rho_\mathrm{disk}(\vec{r}) \qquad\mbox{for}\qquad \rho_\mathrm{env}(\vec{r}) 
\lse \rho_\mathrm{disk}(\vec{r}) \hfill{} \nonumber\\
\mbox{and} \\
\hfill{}
\rho(\vec{r}) = & \rho_\mathrm{env}(\vec{r}) \qquad\mbox{for}\qquad \rho_\mathrm{env}(\vec{r}) 
> \rho_\mathrm{disk}(\vec{r}). \hfill{} \nonumber 
\label{eq: env+disk}
\end{eqnarray} 
This assumption ensures a smooth transition from disk to envelope. In this simple 
envelope $\mathrm{+}$ active disk model we do not implement bipolar cavities although there are hints that 
such cavities, which are caused by collimated outflows, i.e. jets, generally exist in YSOs (Edwards 
et al.~\cite{edwards}). Actually, the number of free modeling parameters
should be as small as possible to ensure that we do not overdetermine our approach. Furthermore,
based on images of YSOs at a similar evolutionary state as RY\,Tau, 
Eisner et al.~(\cite{eisner}) mentioned that cavities have effects on the structure of scattered 
light emission, which has its maximum in the NIR wavelength range. 

\begin{figure}[!htp]
\resizebox{\hsize}{!}{\includegraphics[scale=0.52]{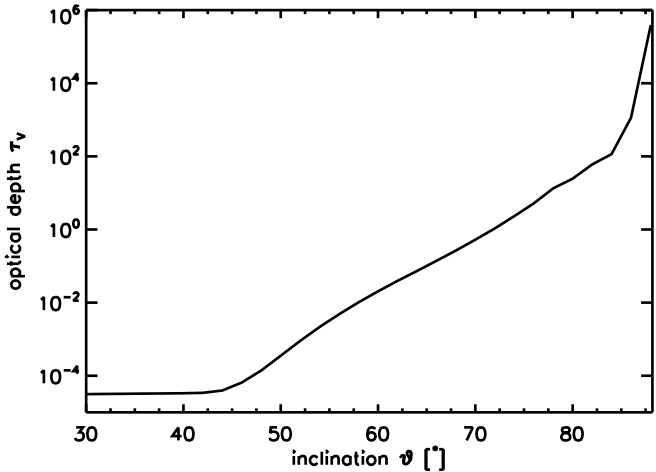}}
\caption{Optical depth for a inclination of $\vartheta$ in the optical
  wavelength range. For smaller inclinations the optical depth is smaller than
  unity ensuring the observations of the inner regions. However, this envelope
  still has effects on the SED and MIR visibilities. }
\label{figure:opdepth}
\end{figure}

\begin{figure*}[!h]
  \centering
  \resizebox{0.3\hsize}{!}{\includegraphics[width=5.3cm]{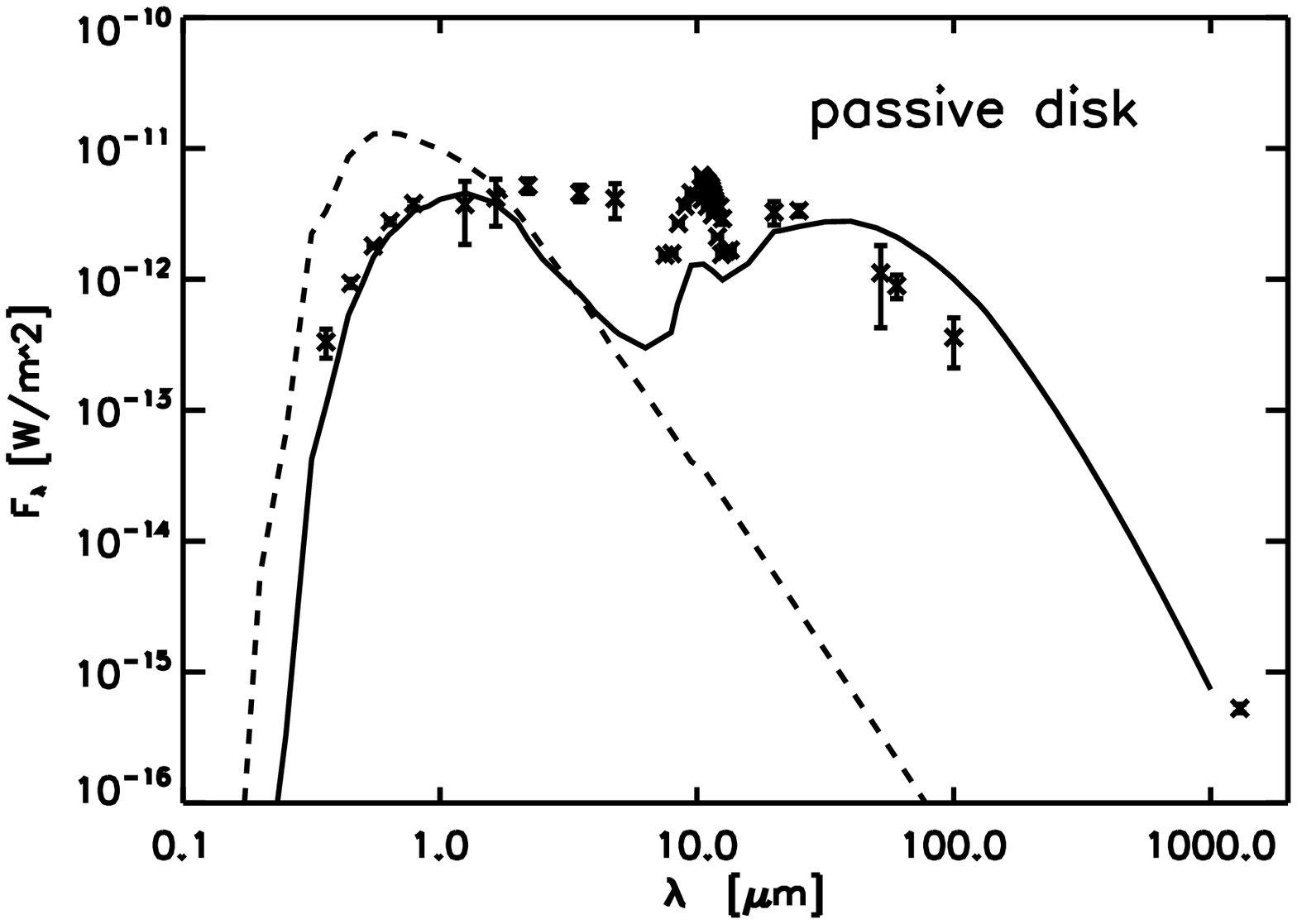}}
  \resizebox{0.3\hsize}{!}{\includegraphics[width=5.3cm]{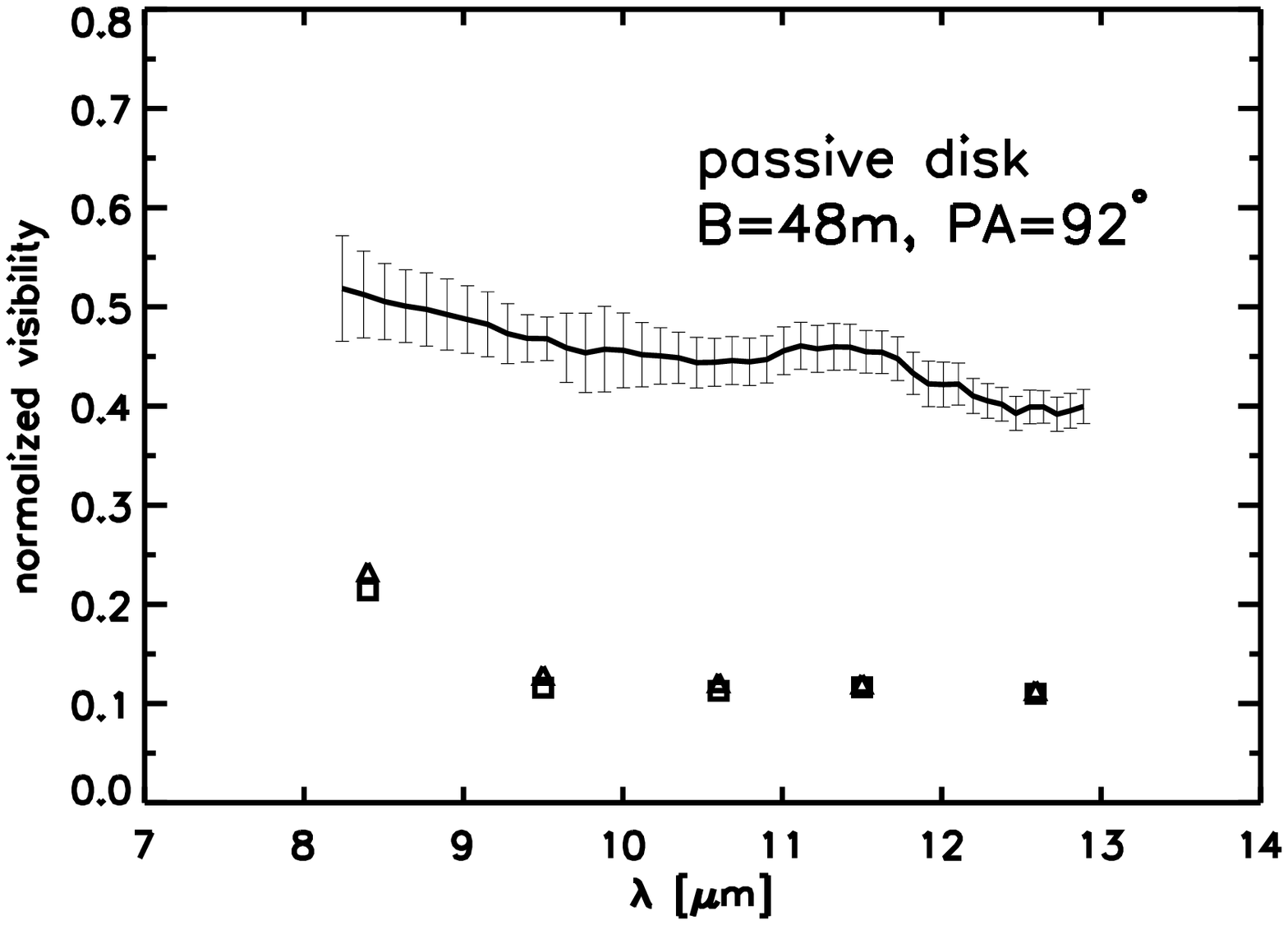}}
  \resizebox{0.3\hsize}{!}{\includegraphics[width=5.3cm]{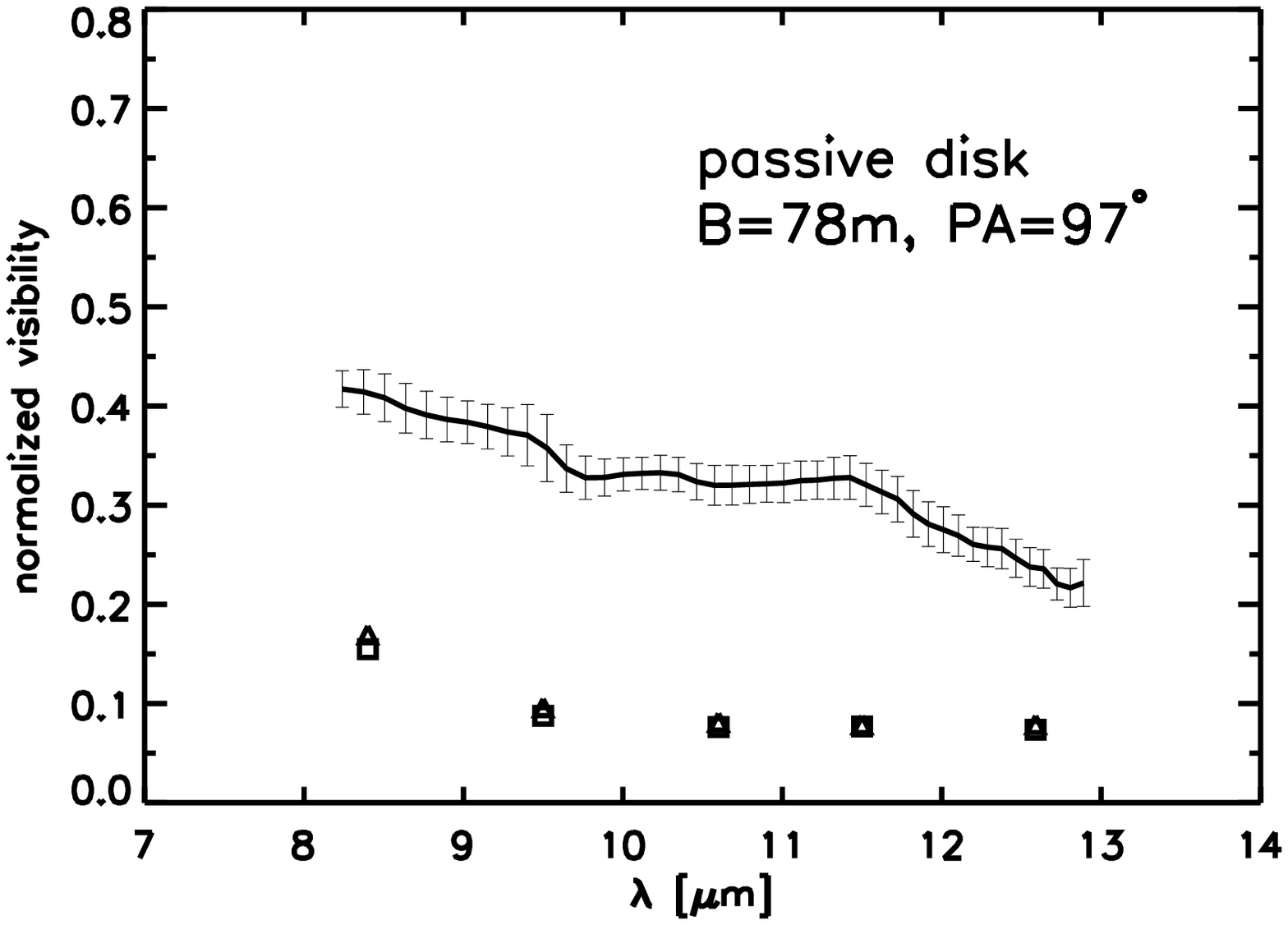}}\newline
  \resizebox{0.3\hsize}{!}{\includegraphics[width=5.3cm]{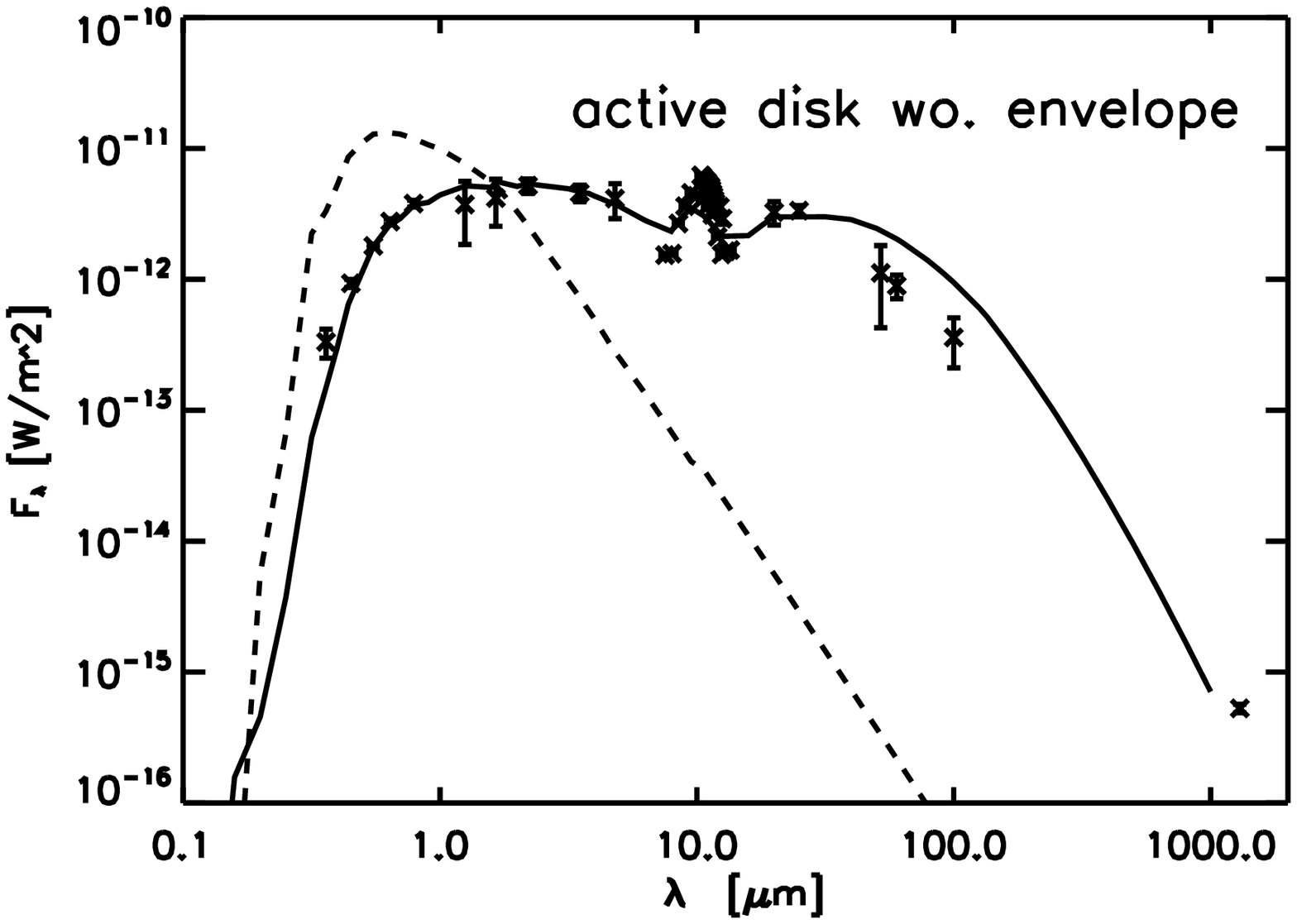}}
  \resizebox{0.3\hsize}{!}{\includegraphics[width=5.3cm]{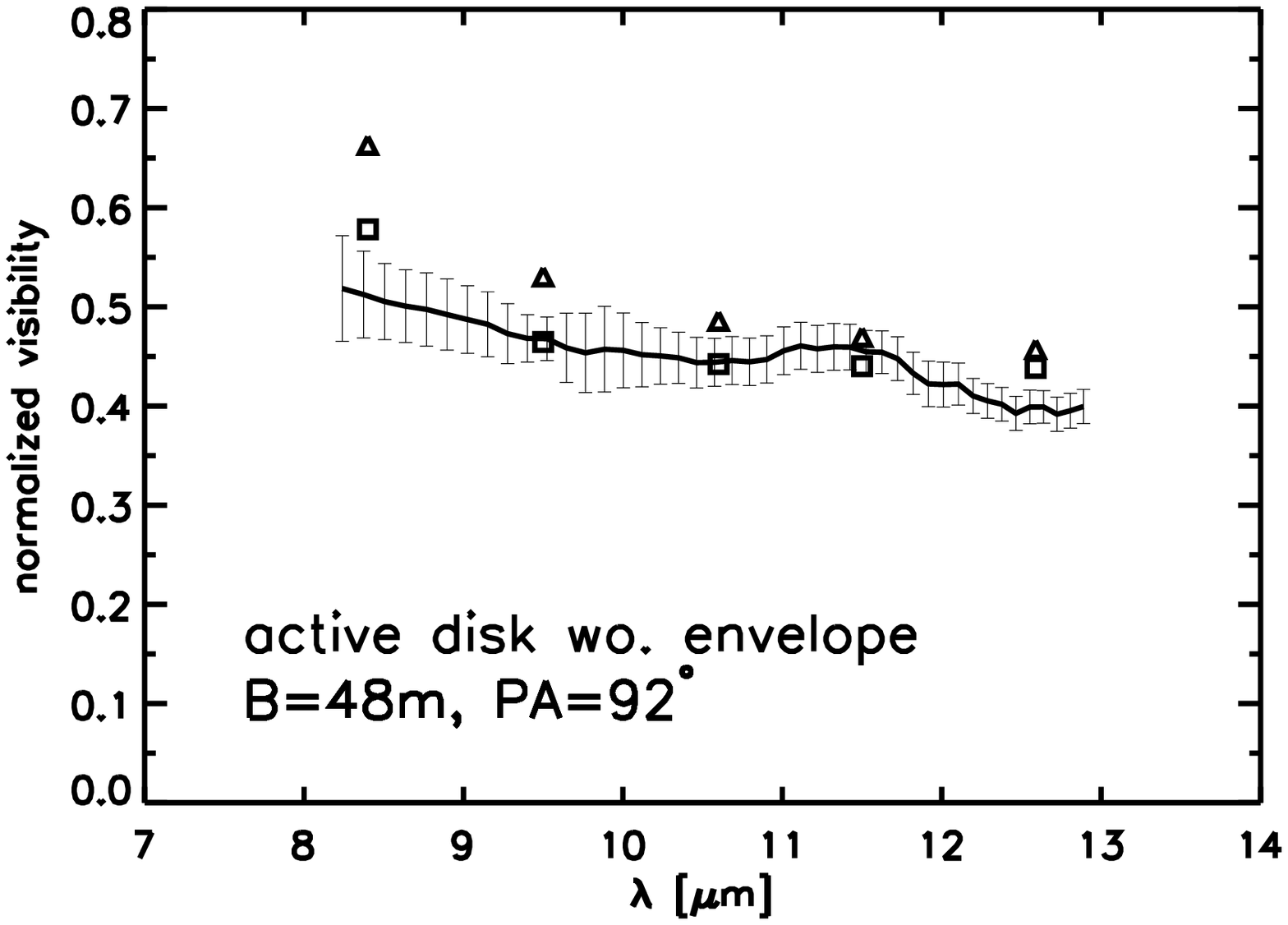}}
  \resizebox{0.3\hsize}{!}{\includegraphics[width=5.3cm]{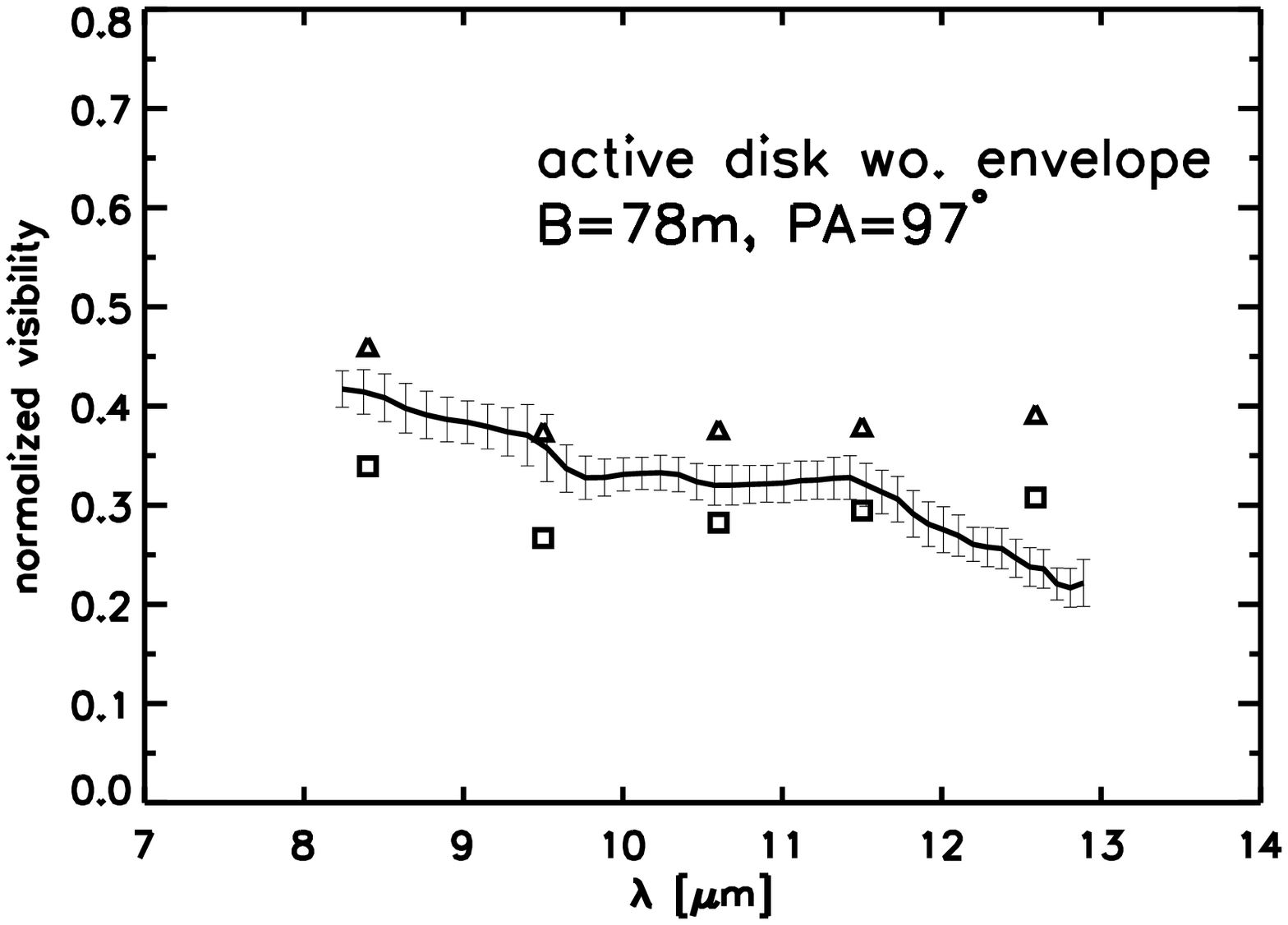}}\newline
  \resizebox{0.3\hsize}{!}{\includegraphics[width=5.3cm]{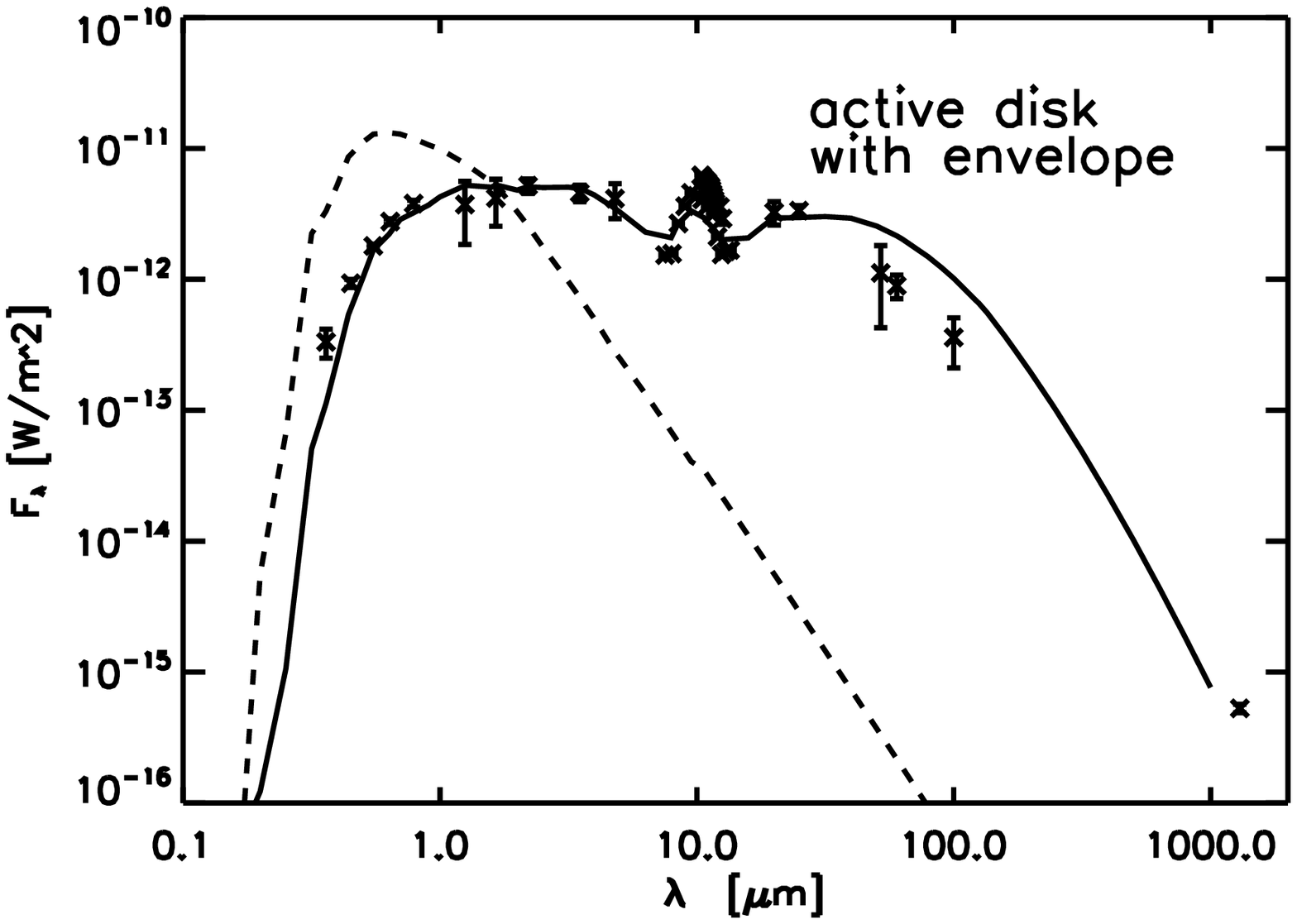}}
  \resizebox{0.3\hsize}{!}{\includegraphics[width=5.3cm]{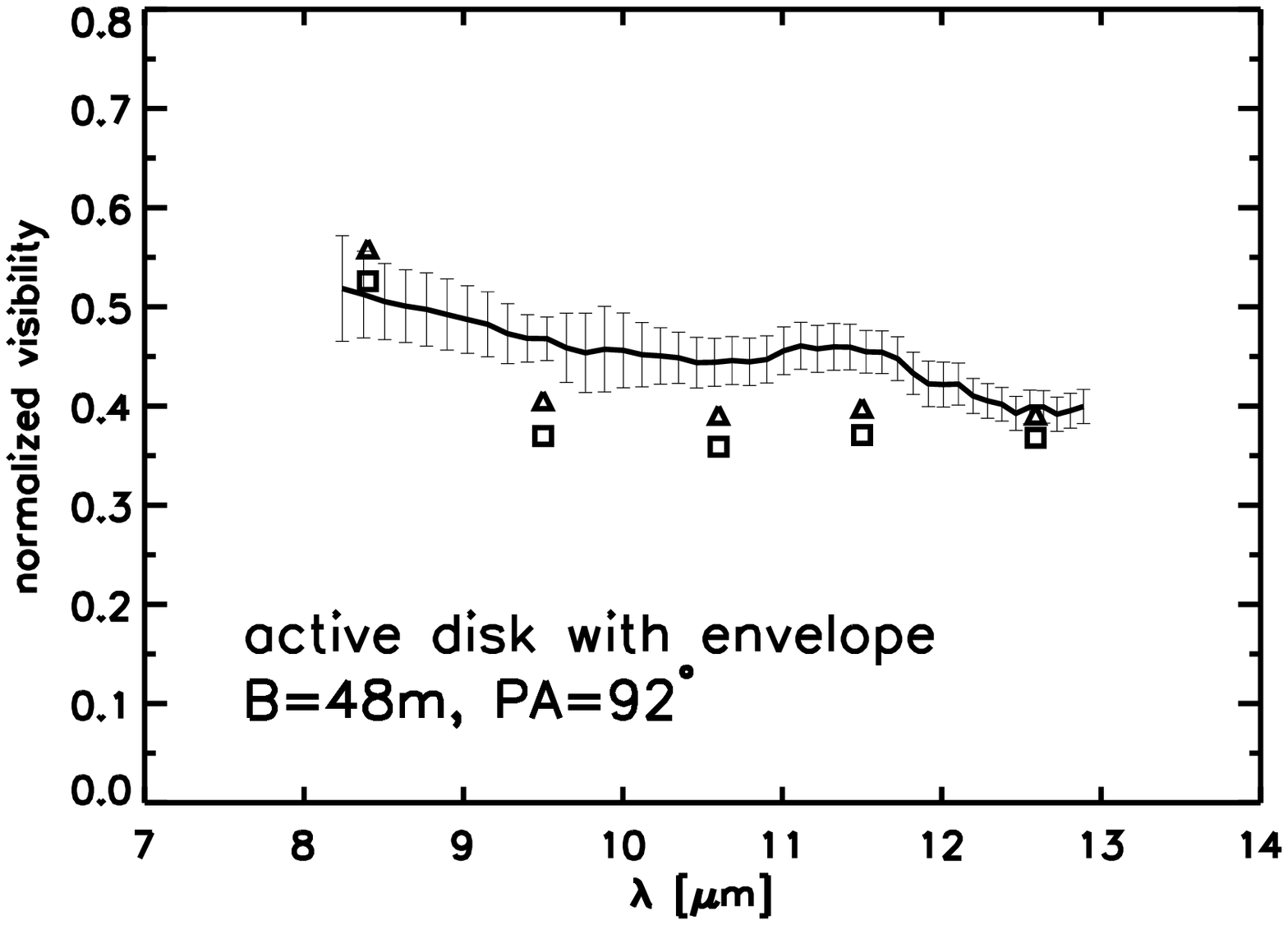}}
  \resizebox{0.3\hsize}{!}{\includegraphics[width=5.3cm]{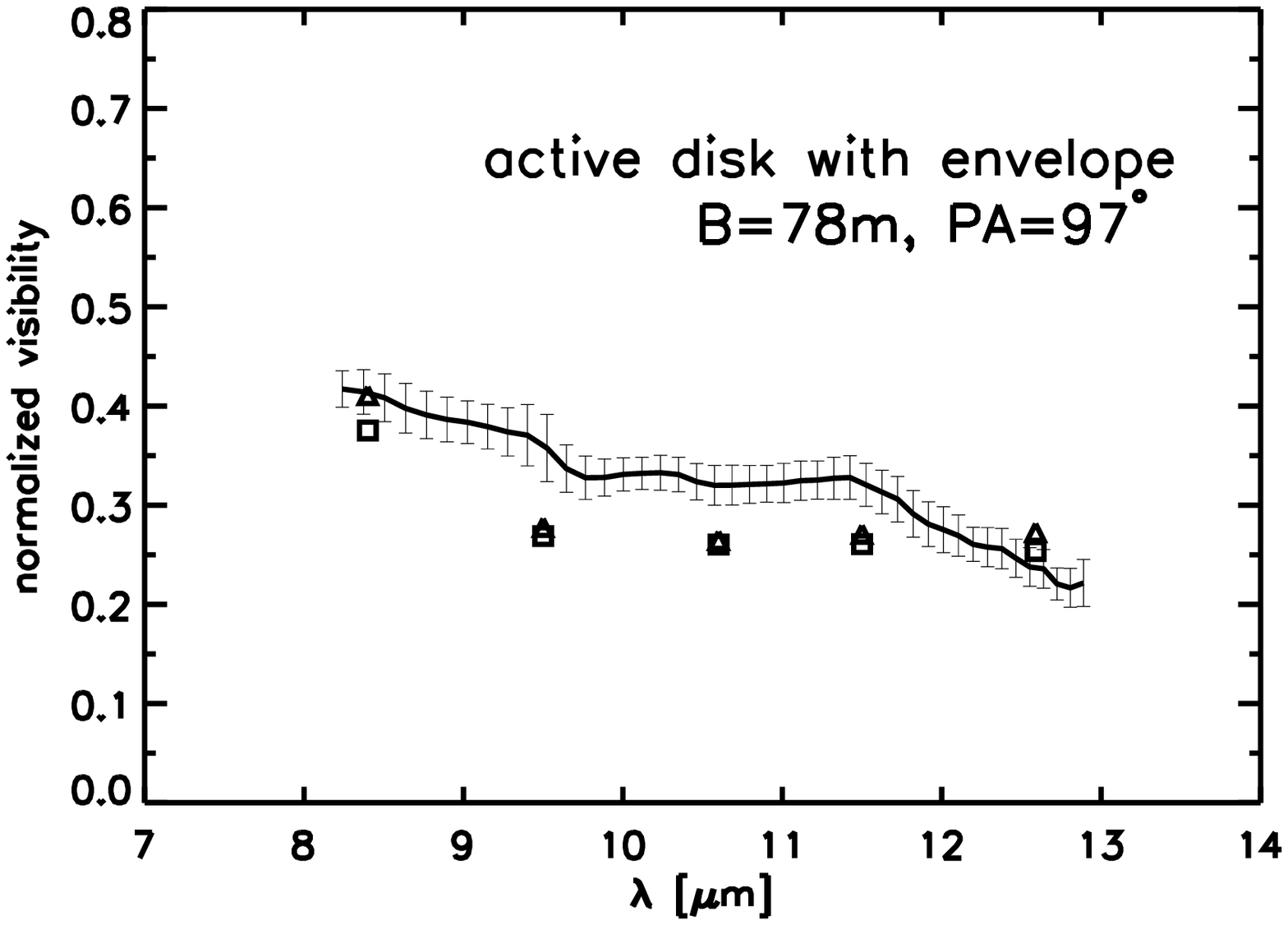}}\newline
  
  \caption{{\it Top row:} Models of the spectral density distribution of RY\,Tau for an 
    inclination angle of $\vartheta=25^{\circ}$. Here, we assume a ``naked'', passively 
    heated disk model without considering accretion effects and an
    envelope. Real
    photometric data with error bars are included in both models (see 
    Table~\ref{table:photometrie}). The dashed line represents the intrinsic, stellar flux. The flux 
    $F_{\lambda}$ is given in units $\mathrm{W m^{-2}}$. According to the theorem 
    of van Cittert-Zernicke the modeled visibilities in all diagrams were calculated from the 
    corresponding model image for the wavelengths of $8.5$, $9.5$, 
    $10.6$, $11.5$ and $12.5\ \mathrm{\mu m}$. Triangles and squares 
    represent the upper and lower limit of the visibilities $V(\lambda)$ for different position 
    angles but the same inclination of the model. The measured data are included with error 
    bars.\newline 
    {\it Middle row:} Model of the spectral density distribution of RY\,Tau for an 
    inclination of $25^{\circ}$ considering accretion effects in addition to a 
    passively heated disk. \newline
    {\it Bottom row:} SED that is obtained only from
    an active disk model considering an envelope, additionally. }
  \label{figure:ry}
\end{figure*}

The Figure~\ref{figure:ry} and Table~\ref{table:ry} show the result and parameter set of our best 
envelope $\mathrm{+}$ active disk models. The accretion rate of the model $\dot{M}=2.5
\times 10^{-8}\ \mathrm{M_\mathrm{\odot}yr^{-1}}$ is by a factor of $\sim$$4$ smaller than the accretion
rate assumed in the previous model without envelope. However, both models
reproduce the SED and the visibilities. This result shows that both accretion and an envelope, 
have the
same effects on the SED and visibilities in the NIR and MIR wavelength range. Moreover, we
have to mention that the measurements could be reproduced without considering
any accretion effects. A comparison of
these results follows in Sect.~\ref{section:discussion}. The accretion
luminosity is $0.3\ \mathrm{L_\mathrm{\odot}}$. In the visual range and for
inclinations $\vartheta \lse 65^{\circ}$ the model is optically thin as can be
seen in Fig.~\ref{figure:opdepth}. However the envelope
evokes an observational effect on the SED and MIR-visibilities. 

\begin{table*}[!h]
  \caption{Parameter set for the passive disk model, the active model
    (Sect.~\ref{section:  accretion}) and the active model with an envelope 
    (Sect.~\ref{section: dusty envelope}). Results correspond to Fig.~\ref{figure:ry}.}
  \centering
  \begin{tabular}{lrrr}\hline\hline
    parameter & passive disk & active disk & active disk $\mathrm{+}$ envelope \\ 
    & model & model & model \\ \hline
    Stellar mass $M_{\star}$ & $1.69\ \mathrm{M_{\odot}}$ & $1.69\ \mathrm{M_{\odot}}$ & $1.69\ \mathrm{M_{\odot}}$ \\
    Stellar temperature $T_{\star}$ & $5560\ \mathrm{K}$ & $5560\ \mathrm{K}$ & $5560\ \mathrm{K}$ \\
    Stellar luminosity $L_{\star}$ & $10.0\ \mathrm{L_{\odot}}$ & $10.0\ \mathrm{L_{\odot}}$ & $10.0\ \mathrm{L_{\odot}}$ \\
    Disk mass $M_\mathrm{disk}$ & $1 \times 10^{-2}\ \mathrm{M_{\odot}}$ & 
    $4 \times 10^{-3}\ \mathrm{M_{\odot}}$ & $4 \times 10^{-3}\ \mathrm{M_{\odot}}$ \\
    Outer disk radius $R_\mathrm{out}$  & $270\ \mathrm{AU}$ & $270\ \mathrm{AU}$ & 
    $270\ \mathrm{AU}$ \\
    Inner disk radius $R_\mathrm{in}$ & $0.3\ \mathrm{AU}$ & $0.3\ \mathrm{AU}$ & $0.3\ \mathrm{AU}$  \\
    Exponent $p$ (see Eq.~\ref{eq: sigma}) & $1.3$ & $1.3$ & $1.4$ \\
    Inclination $\vartheta$ & $\lse 70^{\circ}$ & $\lse 70^{\circ}$ & $\lse 65^{\circ}$ \\
    $c_\mathrm{1}$ (see Eq.~\ref{eq: envelope}) & -- & -- & $5.0 \times 10^{-5}$ \\
    $c_\mathrm{2}$ (see Eq.~\ref{eq: envelope}) & -- & -- & $1.0$ \\
    Accretion rate $\dot{M}$ & -- & $9.1 \times 10^{-8}$M$_\mathrm{\odot}$yr$^{-1}$ & $2.5 \times 10^{-8}$M$_\mathrm{\odot}$yr$^{-1}$ \\
    Boundary temperature $T_\mathrm{bnd}$ & -- & $8000\ \mathrm{K}$ & $8000\ \mathrm{K}$ \\
    Truncation radius $R_\mathrm{bnd}$ & -- & $5\ \mathrm{R_{\star}}$ & $5\ \mathrm{R_{\star}}$ \\  
    \hline
  \end{tabular}
  \label{table:ry}
\end{table*}

\section{Radial gradients of the dust composition in circumstellar disks}\label{section:dust 
  composition}
As mentioned in Schegerer et al.~(\cite{schegerer}) a more advanced disk model considers the 
specific dust composition of the corresponding object instead of using a
canonical MRN dust set with 
averaged optical quantities (see Sect.~\ref{section:dust}). In such dust models the following dust 
components are generally taken into account: carbon, which mainly contributes to the 
underlying continuum as well as amorphous and crystalline 
silicate dust which generates the silicate features at $\sim$$10\ \mathrm{\mu m}$ and $\sim$$20\ 
\mathrm{\mu m}$. Other dust species such as water ice or, more precisely, water ice mantles around 
carbon/silicate grains can also be considered (see Fig.~1 in Chiang et al.~\cite{chiangII}). 

First comparisons between the silicate feature measured in laboratory experiments and 
observationally-based silicate spectra of YSOs have been drawn by J\"ager et 
al.~(\cite{jaeger}). A widely accepted analysis to determine the silicate composition of 
circumstellar dust is a $\chi^2$-fitting method that was established by Bouwman et 
al.~(\cite{bouwman}). They assumed that the silicate emission feature has its origin in the 
optically thin surface layer of the circumstellar disk where it results from a
linear combination of 
mass absorption coefficient (emissivity) $\kappa_{i}$ of different dust components $i$:
\begin{eqnarray}
F(\nu)=B(\nu,T) \left(C_{0} + \sum_{i=1}^{n} C_{i} \kappa_{i}(\nu) \right), 
\label{eq:fit}
\end{eqnarray}
where $C_{0}$ and $C_{i}$ are fitting parameters which reflect the mass
contribution of each component $i$. The quantity $F(\nu)$ is the spectral flux at 
frequency $\nu$, $\kappa_{i}(\nu)$ represents the frequency-dependent mass absorption coefficient for dust component $i$ and 
$B_{\nu}(T)$ is the Planck function corresponding to a blackbody temperature $T$. As basic dust 
set for their $\chi^2$-fitting routine for T\,Tauri objects Schegerer et al.~(\cite{schegerer}) used 
the following silicate species: small ($\mathrm{0.1 \ \mu m}$) and large ($\mathrm{1.5 \ \mu m}$) 
grains of amorphous olivine, and pyroxene, as well as crystalline species such
as forsterite, enstatite, and quartz. 

The single-dish spectrum $F_{\nu}$ is measured with a single telescope. For each interferometric 
observation of RY\,Tau we obtained a correlated spectrum (s. Eq.~\ref{eq: visibility}) which 
reflects the flux emitted by a region which was not spatially resolved by the interferometer. An increasing 
effective baseline length of the interfero\-meter results in a higher resolution. It has to be 
pointed out, that the single-dish spectra as well as the correlated spectra contain spectral 
contributions of the silicate emission from the whole disk, but the contributions from the hotter and 
brighter regions are increasing with increasing effective baseline
length. In this context a homogeneous, axial-symmetric disk is assumed.  

\begin{figure}[!htp]
\resizebox{\hsize}{!}{\includegraphics[scale=0.32]{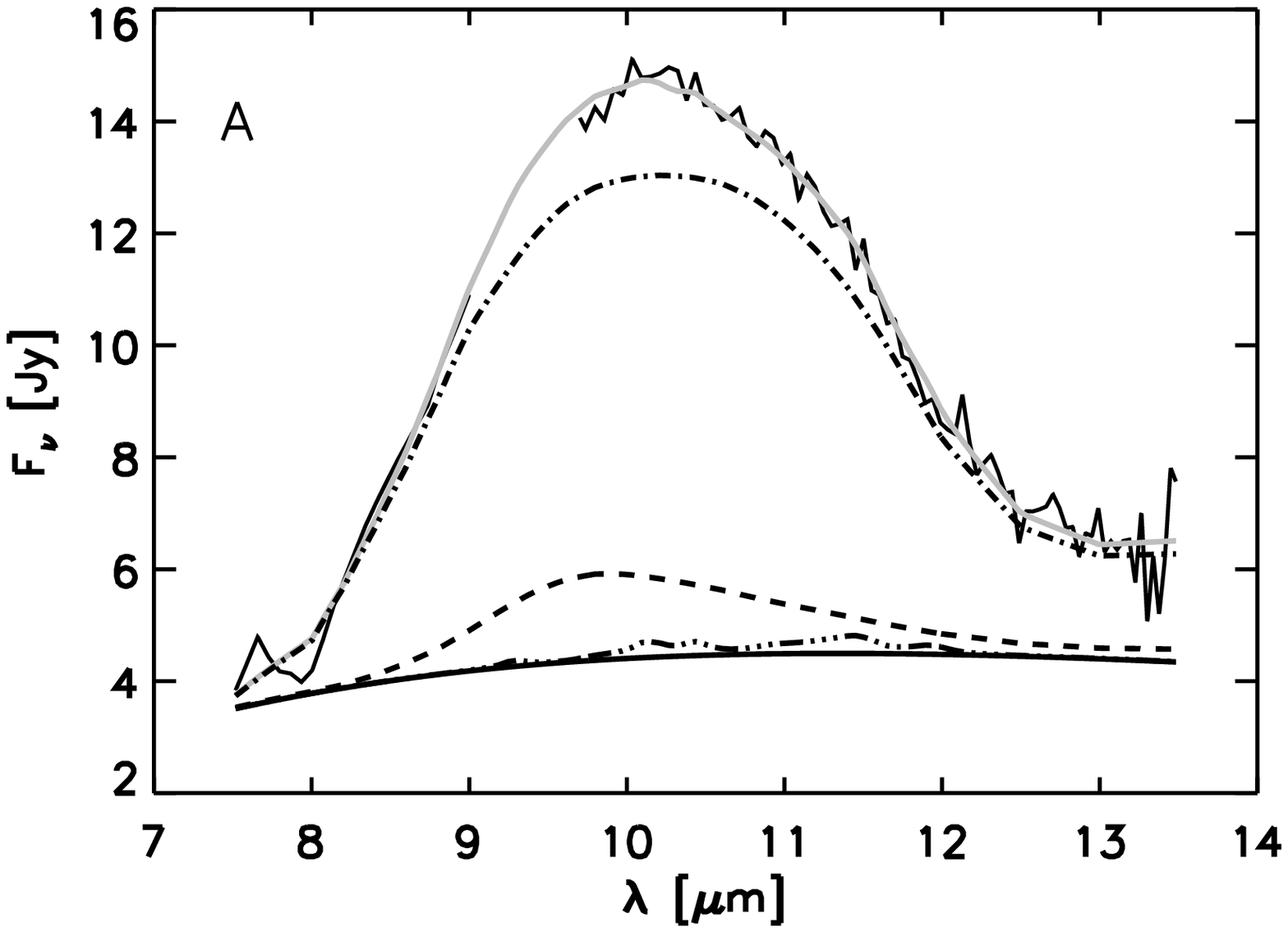}}\newline
\resizebox{\hsize}{!}{\includegraphics[scale=0.32]{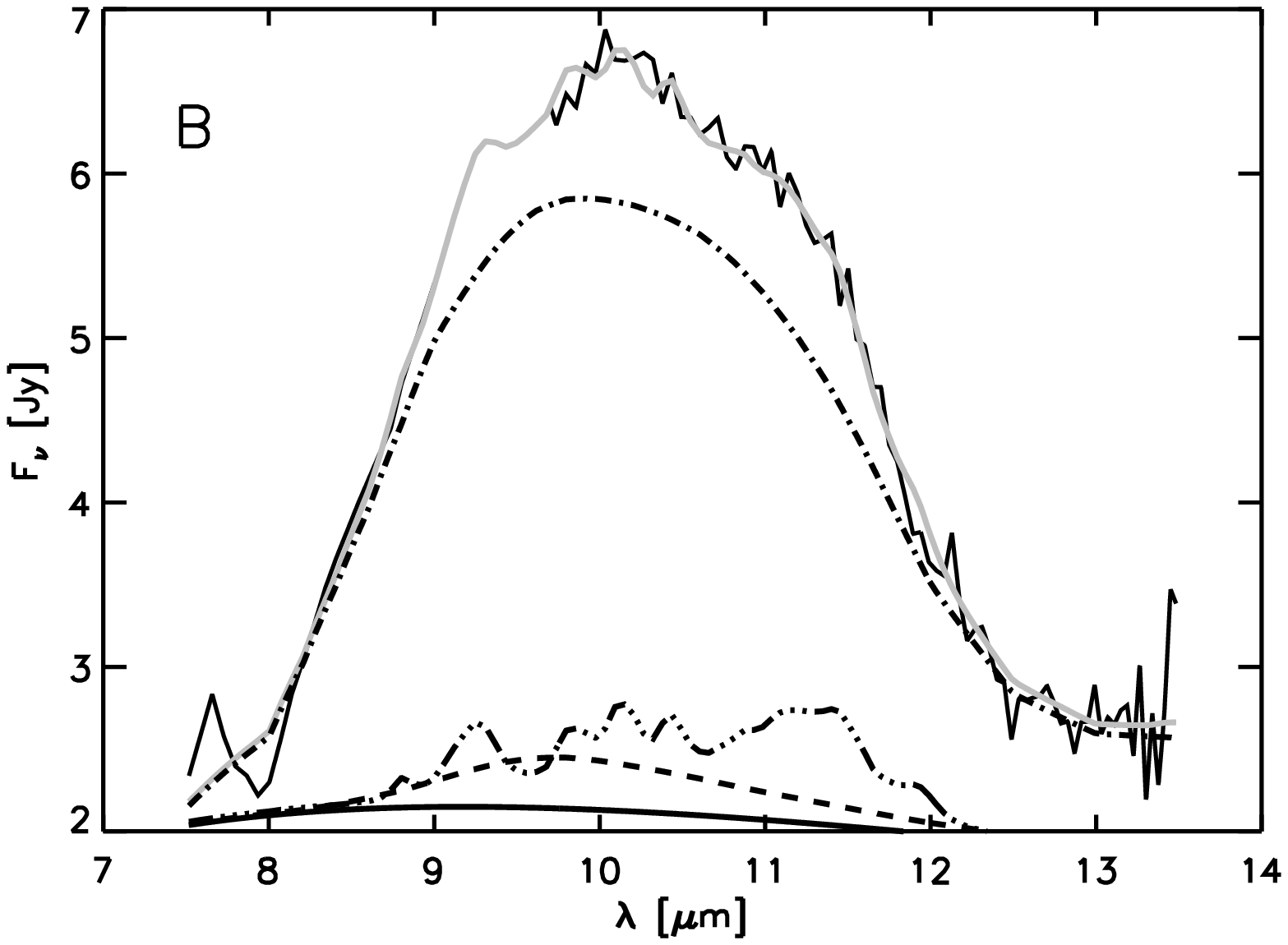}}\newline
\resizebox{\hsize}{!}{\includegraphics[scale=0.32]{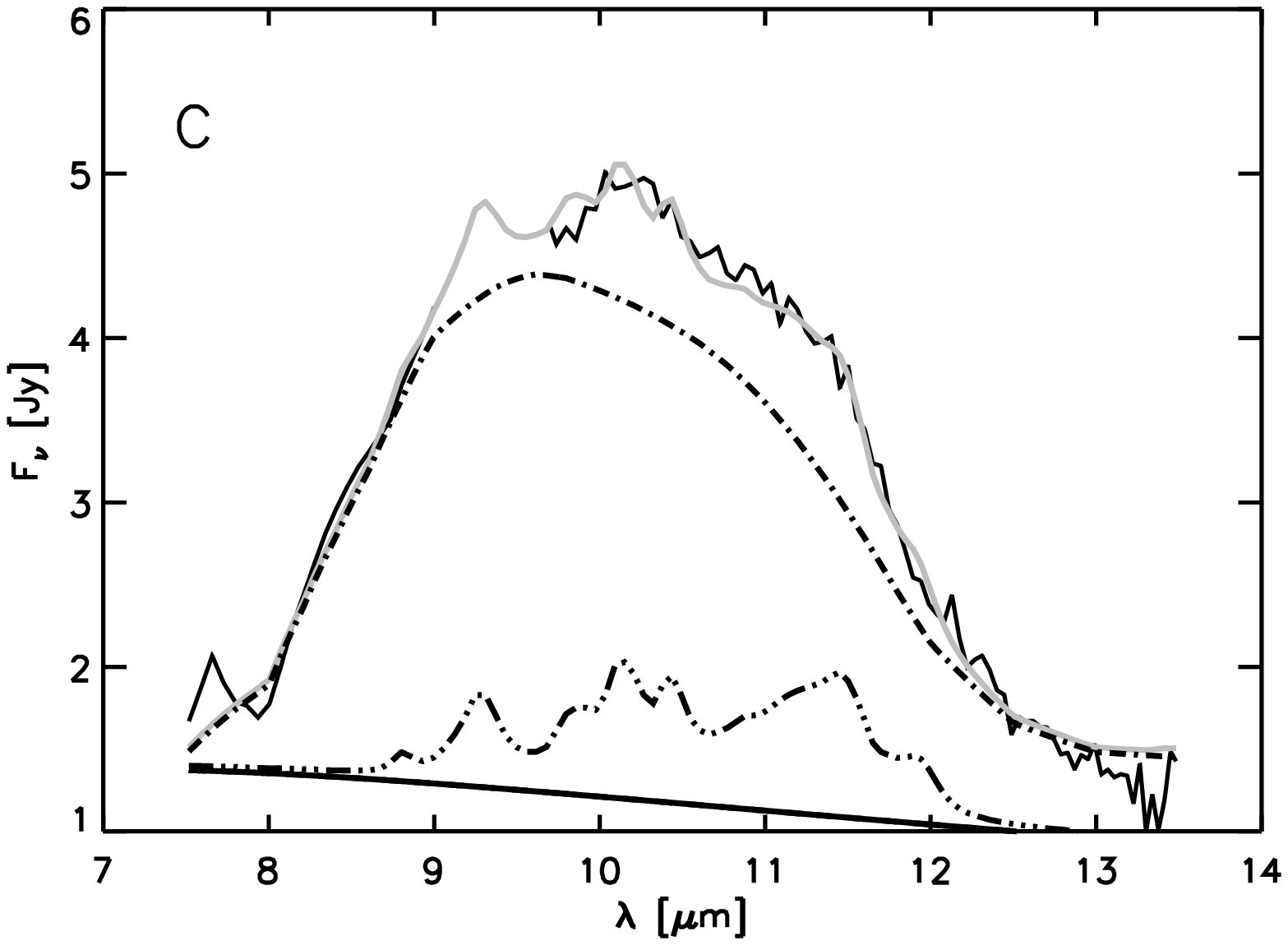}}
\caption{Single-dish (A) and correlated spectra (B) and (C) of 
  the T\,Tauri star RY\,Tau (black, solid lines). The correlated spectra (B) and (C) corresponds to a baseline length of 
  $49\ \mathrm{m}$ and $78\ \mathrm{m}$, respectively. The silicate feature are modeled by a linear 
  combination of mass absorption coefficients $\kappa_{i}$ of different amorphous and crystalline 
  silicates (grey lines). In order to exclude remnants of the data reduction we cut the wavelength 
  interval $\sim$$9.0\ \mathrm{\mu m}$ to $\sim$$9.7\ \mathrm{\mu m}$
  influenced by the telluric ozone band. 
  Dashed lines and dot-dashed lines represent the contribution of $0.1\ \mathrm{\mu m}$- and 
    $1.5\ \mathrm{\mu m}$-sized, amorphous grains, respectively. The dot-dot-dashed curves stand for 
    the crystalline contribution. The underlying, solid curves represent the
    continuum (here: blackbody with temperature $T$). }
\label{figure:fitting}
\end{figure}

According to the method used in Schegerer et al.~(\cite{schegerer}), we find a decreasing 
contribution of not-evolved, i.e. amorphous, $0.1\ \mathrm{\mu m}-small$ dust
grains and an increasing crystallinity with increasing baseline length, 
i.e. decreasing distance to the central T\,Tauri star (see Fig.~\ref{figure:fitting}, 
Table~\ref{table:fitting}). For comparison we add the single-dish, i.e. non-correlated
spectrum $F_{\nu}$ in Fig.~\ref{figure:fitting}. This single-dish spectrum confirms the derived 
tendencies.

\begin{table}[!tb]
  \caption{Results of our $\chi^2$-fit presented in Fig.~\ref{figure:fitting}. The used method is 
    described in Schegerer et al.~(\cite{schegerer}) in detail. The underlying conti\-nuum is 
    estimated by a single blackbody function with the temperature $T$. ``RMC'' stands for 
    relative mass contribution, ``am.--sma.'' for amorphous, $0.1\ \mathrm{\mu m}$-small, 
    ``am.--la.'' for amorphous, $1.5\ \mathrm{\mu m}$-large and ``crys.'' for
    crystalline silicate dust grains. 
    The crystalline component includes small and large silicate species: forsterite, enstatite, and 
    quartz.  See text for further discussion.}
  \centering   
  \begin{tabular}{lcccr}\hline\hline
     & baseline length & resolution & silicate & RMC \\ \hline
     & & & am.--sma. & $(15 \pm 3)\%$ \\
     \raisebox{-1.5ex}[.5ex][.5ex]{A} &\raisebox{-1.5ex}[.5ex][.5ex]{single-dish}& 
     $70\ \mathrm{AU}$ & am.--la. & $(82 \pm 3)\%$ \\
     & & ($520\ \mathrm{mas}$) & crys. & $(3 \pm 1)\%$ \\
     & & & $T$ & $(450 \pm 3)\ \mathrm{K}$ \\ \hline
     & & & am.--sma & $(7 \pm 3)\%$ \\
     \raisebox{-1.5ex}[.5ex][.5ex]{B} & \raisebox{-1.5ex}[.5ex][.5ex]{$49\ \mathrm{m}$} & 
     $2.8\ \mathrm{AU}$ & am.--la. & $(82 \pm 5)\%$ \\
     & & ($21\ \mathrm{mas}$) & crys. & $(11 \pm 2)\%$ \\
     & & & $T$ & $(557 \pm 5)\ \mathrm{K}$ \\ \hline
     & & & am.--sma & $(1 \pm 1)\%$ \\
     \raisebox{-1.5ex}[.5ex][.5ex]{C} & \raisebox{-1.5ex}[.5ex][.5ex]{$78\ \mathrm{m}$} & 
     $1.8\ \mathrm{AU}$ & am.--la. & $(80 \pm 4)\%$ \\
     & & ($13\ \mathrm{mas}$) & crys. & $(19 \pm 2)\%$ \\
     & & & $T$ & $(722 \pm 9)\ \mathrm{K}$ \\ \hline
    \hline
  \end{tabular}
  \label{table:fitting}
\end{table}

\begin{figure}[!tb]
\resizebox{\hsize}{!}{\includegraphics[scale=0.5]{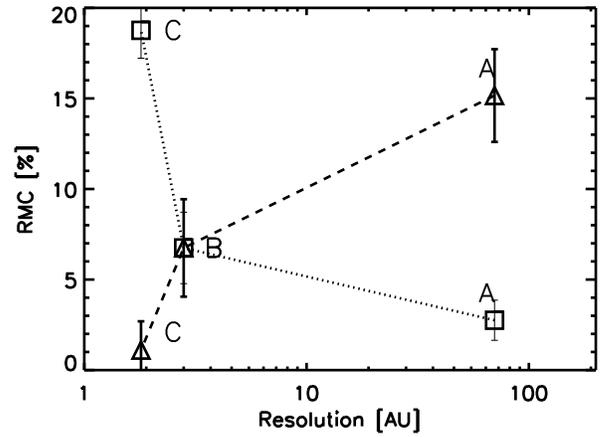}}
\caption{RMC of crystalized (dotted line, squares) and $0,1\ 
  \mathrm{\mu m}$-small, amorphous (dashed line, triangles) silicate grains plotted versus the 
  reached spatial resolution of our observations.}
\label{figure:kees}
\end{figure}

Figure~\ref{figure:kees} shows the crystallinity $C_\mathrm{crys}$, which is plotted 
versus the spatial resolution of our MIDI observations. The crystalized material is concentrated mainly in the 
inner parts of the disk (point C for highest resolution), decreases strongly with decreasing 
resolution (point B for the intermediate resolution) and converge to a lower limit for the 
single-dish observation (point A) which corresponds approximately to the abundance of crystalline dust in 
interstellar matter (Gail~\cite{gailII}). The relative mass contribution of small dust grains 
decreases from the outer to the inner disk regions. Considering the used spectroscopic slit we assume 
a resolution of $0.52\ \arcsec$ for the single-dish observations. A corresponding result was 
previously found by van Boekel et al.~(\cite{boekelII}) for several HAeBe stars indicating more 
evolved silicate dust towards inner disk regions. Our result shows that the formation of crystalized 
dust grains is also favored in the innermost disk region of T\,Tauri stars. A study of the {\it absolute} disk position of the crystalized dust, as for any
other material, is out of the scope of this paper and should be presented in a
future publication. Nonetheless such a forthcoming study is favored by the fact that
the position angles of our observations with MIDI are almost identical
(Table~\ref{table:observation}), i. e. $C_\mathrm{crys}$ depends only on
the radial coordinate $r$ in the disk. Such a study would be an essential
requirement to study the degree of radial mixing of material in circumstellar
disks described by the ratio of the viscous inward and diffusive outwards
stream (Wehrstedt \& Gail~\cite{wehrstedt}; Bouwman et al.~\cite{bouwmanII};
Gail~\cite{gailIII}; Pavlyuchenko \& Dullemond~\cite{pavlyuchenko}). Moreover,
such a study could give a hint whether circumstellar dust grains are only
crystalized by thermal heating in the inner disk regions or whether electric
discharges in outer disk regions can also crystalize dust grains. Both, a
strong radial diffusion and electric discharges in outer disk regions,
respectively, would imply a shallow decrease of the crystallinity with
radius. Although Fig.~\ref{figure:kees} qualitatively shows that crystalized dust grains
are located mainly in the inner disk region, a further study is required to 
determine the {\it absolute} disk position of the crystallised dust, along with 
more interferometric observations with different spatial resolutions.     

Finally, we add a few comments on the dust set used in our modeling approach.  
Instead of using the canonical dust set of ``astronomical silicate'' and carbon, we could also use the specific silicate dust composition found by the simple linear 
fitting routine presented above. Preibisch et al.~(\cite{preibisch}) have already used such an 
approach for modeling the HAeBe star \object{HR\,5999}. However, this proceeding is 
problematic in light of following arguments: 
\renewcommand{\labelenumi}{\roman{enumi}.}
\begin{enumerate}
\item The best-fit model parameters are not sensitive to very small grains ($< 0.1\ \mathrm{\mu m}$). 
  But such tiny grains are essential for generating the observed NIR emission as Weingartner \& 
  Draine~(\cite{weingartner}) pointed out. As mentioned in Sect.~\ref{section:dust}, a 
  dust set where large dust particles $a > 1\ \mathrm{\mu m}$ are implemented at the expense 
  of tiny dust grains, is very ineffective in generating NIR and MIR flux. Only additional modifications of 
  the inner disk structure can compensate this lack of emission. However, in disk systems where 
  the NIR flux has already vanished, the substition of small grains by large grains could be 
  physically justified by advanced dust coagulation (Weinberger et al.~\cite{weinberger}; Kornet et 
  al.~\cite{kornet}). 
\item The best-fit model parameters are not sensitive to carbon. The emission profile of carbon is 
  strictly monotonic in the $10\ \mathrm{\mu m}$-wavelength range (e.\/g., Wolf \& 
  Hillenbrand~\cite{wolfII}) and thus contributes to the underlying continuum,
  only. If carbon is 
  considered in Eq.~\ref{eq:fit}, too, each potential contribution $C_\mathrm{carbon}$ of carbon 
  would not be independent of the contribution of the single blackbody function used to 
  reproduce the underlying continuum.
\item Apart from the analysis of the contributions of each silicate component, another important 
  result of our $10\ \mathrm{\mu m}$-analysis is the derived tendency of increasing crystallinity 
  towards inner disk regions. However, this result does not enclose the determination of the 
  absolute disk position of the crystallised dust and it would be another requirement to 
  use the results of our $\chi^2$-fitting routine in our modeling approach. 
\item It has  not been clear, so far, if the determined 
  contribution of each dust component corresponds to its mass fraction. As shown in 
  Schegerer et al.~(\cite{schegerer}) the porosity of dust grains, which is not considered in our 
  $\chi^2$-fit, affects the shape of the silicate feature similar to the
  size effect. Therefore,
  the mass contributions 
  of large, compact dust grains could be overestimated while the increasing
  crystallinity of the dust 
  in the inner regions is still a safe conclusion (Sect.~\ref{section:dust composition}).  
  
\end{enumerate}
Only further studies can clarify if and how the results of the presented $\chi^2$-fitting routine 
can be implemented in the modeling approach.

\section{Discussion}\label{section:discussion}
The main aim of this paper was to model the structure of the circumstellar environment of the 
T\,Tauri star RY\,Tau. In this context we presented different modeling approaches for the 
circumstellar dust distribution (passive disk; active disk; active disk
$\mathrm{+}$ envelope) 
and pointed to potential supplements such as the puffed-up inner rim wall or the truncated outer disk 
model. An important aim of our approach was to keep the number of model parameters as small as 
possible. With respect to our modeling results additional parameters and modifications were only 
implemented if significant improvements could be obtained afterwards. In order to decrease the mm 
slope of the resulting SED, the disk with the MRN dust composition of 
astronomical silicate and carbon were replaced by a two-layer dust model where the disk interior 
also contains evolved, i.e. larger dust grains.

\subsection{The merits of the active disk model with and without an envelope.}
Accretion or dusty envelopes produce the additional infrared flux from
$\lambda \approx 2\ \mathrm{\mu m}$ up to $\lambda \approx 8\ \mathrm{\mu m}$
of the SED that is missed in the model of a ``naked'', passively heated disk. 

The extra infrared radiation from the active disk model is generated in the
innermost disk  
region inwards the point of dust sublimation (Appendix~\ref{appendix}). The radiation 
which is caused by accretion additionally heats the innermost disk regions close to the inner disk 
radius. In comparison to a ``naked'', passively heated disk, the implementation of accretion effects 
in the model results in a higher spatial concentration of infrared flux in the inner region 
that is not spatially resolved by our MIDI measurement. The computed MIR vi\-sibility 
increases, therefore (compare middle and lower panel in Fig.~\ref{figure:ry}). The accretion rate of 
the model presented here (active disk without envelope), $9.1 \times
10^{-8}$M$_{\odot}$yr$^{-1}$, is smaller than the value found by  
Akeson et al.~(\cite{akeson}; $2.5 \times 10^{-7}$M$_{\odot}$yr$^{-1}$) in
their model study but
corresponds to that of Calvet et al.~(\cite{calvetII}; $6.4-9.1 \times 10^{-8}$M$_{\odot}$yr$^{-1}$). The 
latter result is based on a multi-wavelength study in the optical-UV range considering different 
emission-line profiles. An accretion rate of $2.5 \times
10^{-7}$M$_{\odot}$yr$^{-1}$ decreases the visibilities in the MIR range because of
a stronger irradiation of outer disk regions. Based on a flux ratio measurement between the
continuum excess and the intrinsic photospheric 
flux at a wavelength of $5700\ $\AA, Hartigan et al.~(\cite{hartiganII}) found a much lower accretion 
rate of $\dot{M} \approx 2.5 \times 10^{-8}\ \mathrm{M_{\odot} yr^{-1}}$ in RY\,Tau. Vink et 
al.~(\cite{vink}) determined an accretion rate of $\dot{M} \approx 7.5 \times 10^{-8}\ 
\mathrm{M_{\odot} yr^{-1}}$. The derived accretion rate from the model of Akeson et 
al.~\cite{akeson} is up to a factor of $10$ larger than the values that were 
measured in the UV range. Such a discrepancy between standard accretion disk models and the 
measurements has been already discussed by Muzerolle et al.~(\cite{muzerolle}). To obtain a 
consistency in the measurements, they introduced an artificially puffed-up inner rim in their 
modeling approach, accounting for the large NIR excesses of classical T\,Tauri stars, but 
without 
requiring excessive accretion rates. Instead of implementing such a puffed-up inner rim we assumed an 
envelope in our active disk $\mathrm{+}$ envelope where we assumed a smaller
accretion rate (factor $\sim$$4$) than in the pure active disk model.

NIR flux in the model with the envelope has its origin close to but outwards of the sublimation 
point. This extra flux results from the stellar heating of the dust in the envelope up to the 
sublimation temperature and in an increase of the MIR visibility corresponding to the 
pure active model. 

In contrast to pure accretion, a dusty envelope dims the central star and prevents the outer 
disk regions from being heated too strongly by direct stellar irradiation. Therefore, the MIR 
reemission from these outer regions is decreased and the spatial concentration of the infrared 
radiation in the inner regions is increased. Such an effect caused by the 
envelope results in an increase of the MIR visibility, in particular for measurements with the 
smaller projected baseline of $48\ \mathrm{m}$. We have to mention that
comparable effects could also be achieved by the truncation of the outer disk in 
our active disk model or by a strongly puffed-up inner rim.  

In contrast to our finding Akeson et al.~(\cite{akeson}) did not find any hints for an additional 
envelope in their modeling approach for RY\,Tau  based on NIR, interferometric 
observations. They suggest NIR unveiled CO absorption lines which RY\,Tau is exhibiting. In fact, 
such unveiled CO lines could be evidence of the absence of a substantial envelope (Najita et 
al.~\cite{najita}; Calvet et al.~\cite{calvet}). However, the almost unveiled CO absorption 
lines do not necessarily exclude an envelope when considering the following argument: 

The results of two-dimensional collapse calculations of the infalling matter in an envelope (Yorke et 
al.~\cite{yorke}) suggests a more plane-parallel than spherical envelope geometry.  In this model, the 
envelope has already collapsed at the inner disk edge. 
With MIDI we resolved inner disk regions with a distance of several AU from the star (see 
Fig.~\ref{figure:leinert}) where remnants of the envelope could still exist. However, at these 
distances the gas in the potential envelope 
is already too cold to provide a substantial veiling of CO lines. We point to a study of Bastien 
\& Landstreet~(\cite{bastien}) where it was suggested that most of the polarization found towards 
RY\,Tau arises from a circumstellar (dusty) envelope which actually lies outside of the high-temperature, 
gas-emitting region. Therefore, a geometry where the envelope has already disappeared at the inner 
edge but not at adjacent regions could explain why the observed NIR CO absorption lines are unveiled 
and why Akeson et al.~(\cite{akeson}) failed to model photometric and NIR visibility data 
considering an additional envelope to their pure active disk model. In fact,
with respect to Fig.~\ref{figure:leinert} the NIR 
emission mainly originates close to the sublimation point.

Another potential origin of such an (not necessarily spherical)
envelope could be magnetically driven disk winds containing gas {\it and,
  additionally}, small quantities of small dust particles: material at
the disk surface could even follow magnetical field lines for radii $r$
which are much larger than the magnetical truncation radius $r \gg
R_\mathrm{bnd}$ (Appendix~\ref{appendix}) and
larger than the inner radius $r > R_\mathrm{in}$ (Blandford \&
Payne~\cite{blandford}). Former studies assumed a 
correlation between the disk accretion rate $\dot{M}$ onto the star and the
outflow mass-loss rate $\dot{M}_\mathrm{of}$ with $\dot{M}_\mathrm{of}/\dot{M}
\sim 0.1$ (Richer et al.~\cite{richer}). These dust particles that follow the
outflowing wind from the disk form the optically thin dusty envelope assumed
in our modeling approach. A possible consequence of this procedure would be
the (acceleration of the) formation of an inner gap in the innermost disk
region as observed in older T\,Tauri objects such as TW\,Hya (Calvet et
al.~\cite{calvetIII}). Finally, we note that Fendt \& 
  Camenzind~(\cite{fendt}) have studied stationary, axisymmetric wind
  flows driven by a rapidly rotating magnetosphere. They found that the (gas)
  particle density in the outflow decreases with $r^{-2.3}$. This result is
  independent from the stellar parameters (private communication with
  Ch.~Fendt).

Another origin of the dust particles in the circumstellar envelope could be hard UV 
irradiation from the inner accretion zone or/and the star which increases the
gas temperature in the upper layers of the disk up to $\sim$$10^4\ \mathrm{K}$ 
allowing the hot, gaseous material to escape from the gravitationally bound
system of the star. Small dust particles accompany the gas 
outflow. This effect, called photoevaporation, effectively starts at a critical radius 
$r_\mathrm{cr}$ where the sound speed is in the range of the escape speed,
i. e. a few AU in the case of RY\,Tau (see Dullemond et al.~\cite{dullemondIII}
for a review). However, only further theoretical studies can clearify if the
mass loss rates of dusty material in these disk outflows is high enough to
cause a sufficiently strong effect visible with MIDI. For 
completeness we note that gas pressure dependent photophoretic
forces of light can also induce the ejection of dust from the optical thin
surface layers of the disk as studied by Wurm~(\cite{wurm}) and Wurm \& Krauss~(\cite{wurmII}). 

YSOs of Class I (Adams et al.~\cite{adams}) that reveal a circumstellar
envelope $\mathrm{+}$ disk 
structure typically also show signs for accretion given that accretion is stronger for younger objects 
(e.\/g., Hartmann et al.~\cite{hartmannII}). Accretion is certainly present in RY\,Tau, too, 
according to the mentioned, numerous studies, such as the analysis of the
existing Br$\gamma$ and H$alpha$ lines and UV excess radiation. However, we have also noted that the
implementation of accretion can be totally ignored in an
envelope $\mathrm{+}$ {\it passive} disk model to reproduce SED and MIR visibilities. Both, 
accretion and envelope, increase the NIR- and MIR-flux. Only complementary observations in the UV 
range where the accretion rate can be independently measured, will provide additional constraints to 
disentangle the different model approaches and allow us to consider both accretion and the envelope 
in one model. 

\subsection{A potential stellar companion?}
As mentioned in Sect.~\ref{section:previous measurements}, Bertout et al.~(\cite{bertout}) found 
indirect hints for a stellar companion analysing HIPPARCOS data. Assuming the regular motion of 
the photocenter of RY\,Tau they derived a projected minimum distance of $3.27\ \mathrm{AU}$ and a 
position angle of $304^{\circ} \pm 34^{\circ}$ for the potential secondary. The method they used is 
described in Wielen~(\cite{wielen}). 

\begin{figure}[!htp]
\resizebox{.45\textwidth}{!}{\includegraphics{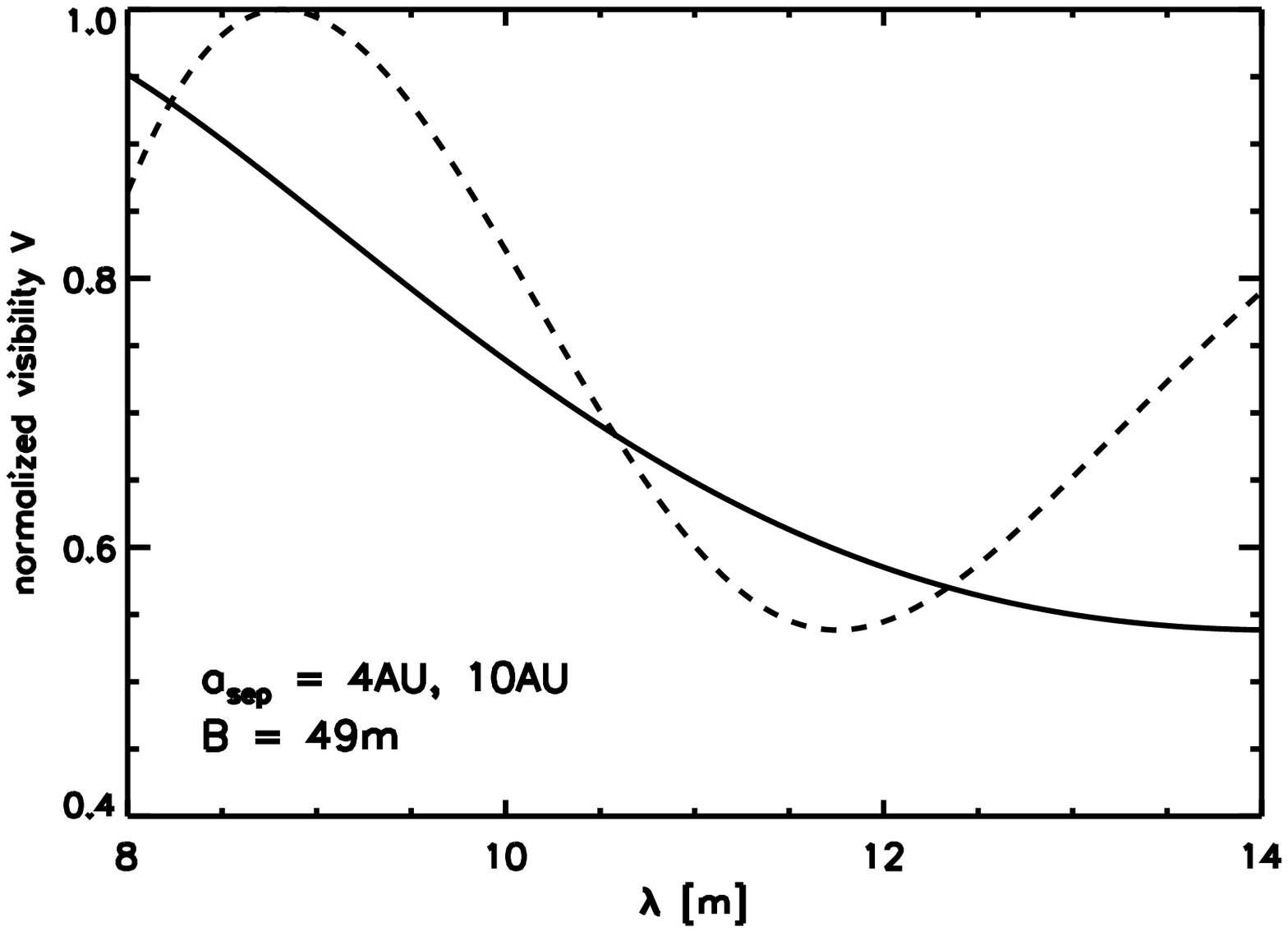}}
\resizebox{.45\textwidth}{!}{\includegraphics{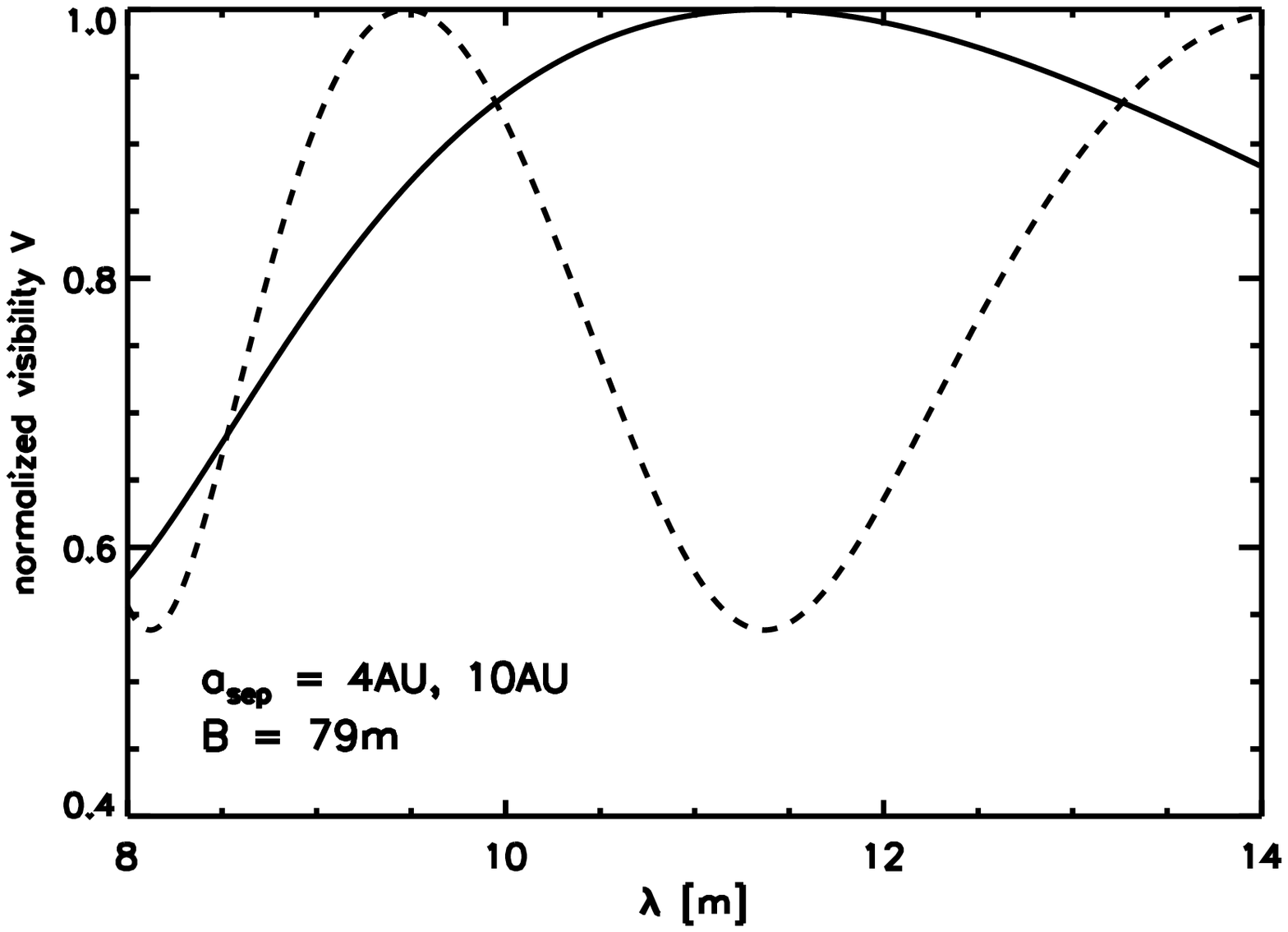}}
\caption{Theoretical prediction for the result obtained for the observation of
  a stellar binary sytem with MIDI. Here, we assume two point sources, i. e. 
  $a_\mathrm{0}(B,\lambda)=1$ in Eq.~\ref{eq:thorsten}. The (normalized)
  visibility is shown in
  relation with the separation of both stars (solid lines: $a_\mathrm{sep}=4\
  \mathrm{AE}$; dashed lines: $a_\mathrm{sep}=10\ \mathrm{AE}$)
  and the projected baseline length (left: $B=49\ \mathrm{m}$; right:
  $B=79\ \mathrm{m}$). The brightness ratio of the components are $1:3$. In
  this example the baseline of the interferometer is also parallel to the
  connection line of the components.}
\label{figure:thorsten}
\end{figure}

A potential detection of a companion by interferometric observations depends
on the separation $a_\mathrm{sep}$, the 
position angle of the companion with respect to the position angle of the interferometric 
baseline and the brightness ratio $a_\mathrm{rat}$ of secondary and primary. 
The visibility of a binary system can be expressed by the approximate formula:
\begin{eqnarray}
\hfill{}
V(B,\lambda)=a_\mathrm{0}(B,\lambda)
    \frac{\sqrt{ 1 + a_\mathrm{rat}^2+2a_\mathrm{rat}\cos(2 \pi a_\mathrm{sep} \frac{B}{\lambda} )}}
         {1+a_\mathrm{rat}}.
\hfill{}
\label{eq:thorsten}
\end{eqnarray} 
The parameter $a_\mathrm{0}(B,\lambda)$ represents the visibility that results
from the circumstellar and -binary material assumed to be equally distributed around the components. 
A positive detection could then be 
recognized by sinusoidal variation of the visibility (Ratzka~\cite{ratzka}). Furthermore, all 
interferometric observations 
are snapshots of the system in contrast to the HIPPARCOS
observations repeated on long-term. A disadvantageous 
configuration of the secondary could actually prevent detection with the interferometer. With the 
non-dection of a clear binary signal in our MIDI data, we can thus neither verify, nor disprove the 
existence of a companion. Assuming $a_\mathrm{0}(B,\lambda)=1$ for two point
sources, Fig.~\ref{figure:thorsten} shows the relation between the
interferometric binary signal and the projected baseline and the separation of
the objects, respectively. With increasing baseline length or increasing
separation the frequency of the sinusoidal variation also increases. The
amplitude of the variation decreases with the brightness ratio of the components. 

Another aspect of this discussion is the possibility that the regular motion of the photocenter 
observed with HIPPARCOS in the visible range is not caused by a secondary, but by a 
brightness irregularity in the circumstellar environment, i.e. in the disk and/or envelope. In fact, 
considering scattered-light images of Class I objects (Padgett et al.~\cite{padgett}), 
circumstellar envelopes, which have their brightness maximum in the NIR and adjacent wavelength ranges 
(Wolf et al.~\cite{wolfIV}), reveal such brightness irregularities. However, it is not clear if 
such a brightness irregularity in the circumstellar environment of RY\,Tau is strong enough to cause the 
observed regular motion of the photocenter.

\subsection{Comparison with HAeBe stars}
HAeBe stars were formerly classified in Group I and Group II sources (Meeus 
et al.~\cite{meeus}; Sect.~\ref{section: inner rim}). Recently, Leinert et al.~(\cite{leinertIV}) 
could explain this phenomenological classification after modeling the interferometric measurements 
of several HAeBe objects obtained with MIDI. In fact, they noticed that the half light radius of 
their disk models in the MIR wavelength range\footnote{The half light radius encircles the disk 
region emitting half of the totally released energy in the disk.} linearly correlates with the 
IRAS color between $12\ \mathrm{\mu m}$ and $25\ \mathrm{\mu m}$, i. e. 
$-2.5 \log(F_\mathrm{\nu}(12\ \mathrm{\mu m})/F_\mathrm{\nu}(25\ \mathrm{\mu
  m}))$. Circumstellar disks around Group II sources are less flared. 
The geometrical effect results in the outer disk regions around Group II sources being less 
strongly heated than the corresponding disk regions of Group I
sources. Therefore, the size of the half light 
radius and, simultaneously, the MIR color is smaller for Group~II sources. 

Because of its stellar luminosity ($L_{\star} = 11.5\ \mathrm{L_{\odot}}$) and stellar mass 
($M_{\star} = 1.69\ \mathrm{M_{\odot}}$) the T\,Tauri star RY\,Tau could be considered as a 
transition object between T\,Tauri and HAeBe stars.  
As the FIR-flux declines, RY\,Tau can be classified as a Group II source (Meeus 
et al.~\cite{meeus}). 
\begin{figure}[!tb]
\resizebox{\hsize}{!}{\includegraphics[scale=0.5]{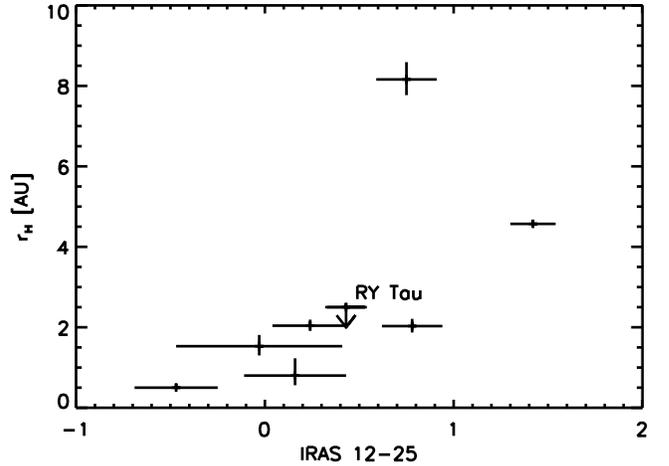}}
\caption{Correlation between the half light radius $r_\mathrm{H}$ at a wavelength of $12.5\
  \mathrm{\mu m}$ and the IRAS-color at $\lambda=12\ \mathrm{\mu m}$ and 
  $\lambda=25\ \mathrm{\mu m}$. Plotted are the corresponding values for HAeBe objects
  (Leinert et al.~\cite{leinertIV}) and for RY\,Tau. The error bars correspond
  to errors of the IRAS fluxes and of the interferometric measurements. The value
  $r_\mathrm{H}$ of RY\,Tau is an upper limit, as the active disk model predicts a too
  small visibility in comparison to the measurement.}
\label{figure:rh-iras}
\end{figure}
The IRAS color of RY\,Tau in the MIR wavelength range is $-2.5 \log(F_\mathrm{\nu}(12\ 
\mathrm{\mu m}) / F_\mathrm{\nu}(25\ \mathrm{\mu m})) = 0,43 \pm 0,11$ (IRAS Catalogs~\cite{iras}). 
The half light radius $r_\mathrm{H}$, i.e. the $FWHM$ of the intensity distribution in our best 
active disk model is determined here by a Gaussian fit. The intensity distribution is 
computed for a wavelength of $\lambda = 12.5\ \mathrm{\mu m}$ as it is not affected by the silicate 
emission at this wavelength. We obtain $FWHM = 1.75\ \mathrm{AU}$. This value
is an upper limit of the half light radius as the active disk model predicts a too
small visibility in comparison to the measurement, i.e. the intensity
distribution of the model disk decreases too slowly for increasing radii in
contrast to the real intensity distribution (measurement). 
However, with respect to the study of Leinert et 
al.~(\cite{leinertIV}, Fig.~5) this result actually confirms the linear correlation formerly found 
for HAeBe stars, solely. A further correspondence between T\,Tauri and HAeBe objects has been already 
shown in Schegerer et al.~(\cite{schegerer}) considering correlations between stellar properties and 
the silicate composition of the circumstellar disk of T\,Tauri and HAeBe objects.

\subsection{Complex interplay of disk parameters}
The visibility, which we try to model, is a complex function of many different disk and dust 
parameters. In such complex disk models the effect of any modifications of 
geometrical disk parameters cannot always be predicted. However, the tendencies of modifications 
of the main model parameters are summarized in the following:

\renewcommand{\labelenumi}{\roman{enumi}.}
\begin{enumerate}
\item An increase of the stellar effective temperature $T_{\star}$ and stellar luminosity 
  $L_{\star}$, in particular, mainly causes an increase of the visible {\it and} infrared flux. 
  If the flux of the unresolved star increases more strongly than the heating of the dust and the MIR 
  flux from the resolved circumstellar surroundings, the MIR-vi\-sibility
  increases. The accretion luminosity  
  affects SED and MIR visibility, correspondingly. An increase of the stellar mass $M_{\star}$ 
  corresponds to an increase of the accretion luminosity in the active disk model ( 
  Eq.~\ref{eq: accretion energy}).
\item The optical depth of the disk increases with the disk mass assuming a constant outer radius. 
  Compacter disk structure decreases the averaged temperature in the inner disk region where less 
  MIR flux is emitted. However, the MIR flux from outer regions simultaneously increases and the MIR 
  visibility therefore decreases.
\item A decrease of the inner disk radius $R_\mathrm{in}$ results in more NIR
  and MIR flux and slightly 
  increases the MIR visibility. 
\item As shown in Sect.~\ref{section: accretion} the outer disk radius determines the FIR 
  flux in our modeling approach. The effect of an increase of the 
  outer radius corresponds to the effect of a decrease of the disk mass.
\item Similar to the outer disk radius, the exponent $p$ has distinct effects on the radial density 
  distribution in the disk. Surface density distributions $\Sigma$ with larger gradients 
  $p$ result in an increase of the density in the inner regions of the disk. The MIR visibility and 
  the infrared excess from the disk in the NIR up to the MIR wavelength range increases, 
  simultaneously. While the FIR excess can decrease strongly at values greater 
  than $p=2.0$, smaller exponents $p<0.8$ only negligibly affect the SED and MIR visibility. 
\end{enumerate}
The spatial information, which we got from the MIDI observations, strongly reduced the number of 
disk models which can reproduce the SED. Certainly, the outer disk radius and the disk mass are less constrained 
parameters in our models. The uniqueness of our approach cannot be
proven. However, in our approach we consistently verified whether the modeling
results can be improved
by varying the modeling parameters. The step sizes used for a final
variation of these modeling parameters are: \medskip\\
\noindent $\Delta R_\mathrm{out}=10\ \mathrm{AU}\mathrm{, }\  
\Delta R_\mathrm{in} = 0.05\ \mathrm{AU}\mathrm{, }\  $ \nonumber\smallskip\\ 
$\Delta L=0.1\ \mathrm{L_\mathrm{\odot}}\mathrm{, }\ \Delta c_\mathrm{2}=0.5\mathrm{, and }\  \Delta p = 0.1.$\medskip\\ 
\noindent For the parameter $c_\mathrm{1}$ and $M_\mathrm{disk}$ we used:\medskip\\
\noindent $\Delta c_\mathrm{1}=0.1c_\mathrm{1},\  
\mathrm{ and }\ 
\Delta M_\mathrm{disk} = 0.5M_\mathrm{disk},\ $\medskip\\
\noindent respectively. These step sizes can be considered as an error of each resulting
parameter (Table~\ref{table:ry}). As mentioned
above, the stellar mass $M_\mathrm{star}$, the effective stellar temperature
$T_\mathrm{\star}$, the accretion rate $\dot{M}$, the boundary temperature
$T_\mathrm{bnd}$ and radius $R_\mathrm{bnd}$ as well as the visual extinction
$A_\mathrm{V}$ are constant parameters considering previous measurements.  

\section{Summary}\label{section:conclusion}
We present interferometric observations of the classical T\,Tauri star RY\,Tau in the $10\ 
\mathrm{\mu m}$ range, which show the source well resolved, together with the total spectrum of the 
source. We modified the MC3D code (Wolf et 
al.~\cite{wolfI}) to obtain a self-consistent model (in the temperature {\it and}
density distribution) of the circumstellar disk around RY\,Tau, 
including accretion, with the following results:
\renewcommand{\labelenumi}{\roman{enumi}.}
\begin{enumerate}
\item Both, the SED and the $10\ \mathrm{\mu m}$ interferometric measurements can be fitted by a 
  self-consistent, physically reasonable radiative transfer model. The
  accretion and an envelope have the effect of filling 
  in the missing emission in the range from $3\ \mathrm{\mu m}$ up to $8\ \mathrm{\mu m}$.
\item A model of a circumstellar active disk with an optically thin envelope
  and a low accretion rate represents the data and the model without
  the envelope and a much larger accretion rate (Akeson
  et al.~\cite{akeson}). The modeling approach with an envelope, which could have its
  origin in an outflowing gaseous {\it and dusty} disk wind, should be considered as
  a future, viable alternative for the puffed-up inner rim, which results from a vertical
  expansion of the disk edge according to Natta et al.~(\cite{natta})
  and Dullemond et al.~(\cite{dullemond}). The latter approach could not be
  confirmed with our particular model of RY\,Tau considering a self-consistently determined density
  distribution considering hydrostatical equilibrium. 
\item The interferometric observation reduces the number of models that solely reproduce 
  the SED of RY\,Tau. 
\item The results of our modeling approaches are: The active disk with or
  without an (optically thin) envelope predicts a system with 
  an upper limit of $\lse 55^{\circ}$ for the inclination. While a stellar
  mass of $M_{\star}=1.69\ \mathrm{M_{\odot}}$ and 
  an effective temperature of $5560\ \mathrm{K}$ can be adopted from previous measurements 
  (Table~\ref{table:properties}), we found a stellar luminosity of $L_{\star} \approx 10.\ 
  \mathrm{L_{\odot}}$ which is smaller than the estimates of Akeson et al.~(\cite{akeson}; $12.8\ 
  \mathrm{L_{\odot}}$) but corresponds to the findings of Calvet et al.~(\cite{calvetII}; $9.6\ 
  \mathrm{L_{\odot}}$). The inner radius of our models $R_\mathrm{in} = 0.3\ 
  \mathrm{AU}$ coincides with the result of Akeson et al.~(\cite{akeson}; $R_\mathrm{in} = 0.27\ 
  \mathrm{AU}$) where interferometrical studies in the NIR preceded. The accretion rate of the model 
  without an envelope is $9.1 \times 10^{-8}$M$_{\odot}$yr$^{-1}$ and corresponds to the upper
  limit found by Calvet et al.~(\cite{calvetII}). The use of an
  additional envelope allows us to use a lower accretion rate of $2.5 \times
  10^{-8}$M$_{\odot}$yr$^{-1}$ which is the same as the result published by Hartigan et 
  al.~(\cite{hartiganII}).
\item Comparison of interferometric and single-dish observations shows for the first time dust 
  evolution in a T\,Tauri star with a reduced fraction of small amorphous and an increased fraction 
  of crystalline particles closer to the star. This is similar to the finding for HAeBe stars (van 
  Boekel et al.~\cite{boekelII}).  
\item Our data neither support nor contradict the existence of a close ($\sim$$3\ \mathrm{AU}$) 
  companion for which some astrometric evidence exists (Bertout et al.~\cite{bertout}).
\item The complex interplay of different disk parameters is consistently apparent from the use 
  of the Monte Carlo radiative transfer code. 
\end{enumerate}

\begin{acknowledgements}
A. A. Schegerer and S. Wolf were supported by the German Research Foundation (DFG) through the 
Emmy-Noether grant WO 857/2 ({\it ``The evolution of circumstellar dust disks to planetary 
systems''}).
\end{acknowledgements}

\appendix
\section{The accretion model}\label{appendix}
Accretion models have been studied in many pu\-blications and the parameters that we used for 
our active disk model are well-known properties of such models. However, 
only a few authors have implemented accretion effects in radiative transfer models (e.\/g., Akeson 
et al.~\cite{akeson}). It is generally assumed that the passive heating of the disk is the 
dominating source of infrared and mm irradiation. This is certainly true for the large scales of the 
disk but not for its innermost region ($< 2\mathrm{AU}$; e.\/g., D'Alessio et al.~\cite{dalessioIII}) 
which can be resolved with long-baseline interferometers. We briefly summarize here 
the parameters of the accretion disk model implemented in our approach.

Former accretion disk models are generally based on the assumption of a geometrically thin, steady 
disk established by Lynden-Bell \& Pringel~(\cite{pringle}) and Pringle~(\cite{pringleII}). 
In this canonical accretion model it is assumed that viscous stresses 
within the disk transport angular momentum to its outer regions. As a consequence of this, most of 
the disk material moves inward onto the protostar, while some disk matter moves outward, 
absorbing all the angular momentum. Assuming a geometrically thin and steady disk the conservation 
of transversal and angular momentum results in the dissipation rate $D$ per unit area and time as a 
function of the radial distance from the star $r$:  
\begin{eqnarray}
\hfill{}
D(r)=\frac{3 G M_{\star} \dot{M}}{4 \pi r^3} 
\left[1-\left(\frac{R_\mathrm{\star}}{r}\right)^{1/2}\right].
\hfill{}
\label{eq: dissipation rate}
\end{eqnarray}
The quantities $G$, $M_{\star}$, $\dot{M}$, $r$ and $R_\mathrm{\star}$ are in
this order: gravity constant, stellar mass, accretion rate of the infalling
material, radial distance and 
stellar radius. The total released energy between an inner boundary radius of the disk 
$R_\mathrm{bnd}$ and infinity is 
\begin{eqnarray}
\hfill{}
L_\mathrm{disk}=\int_{R_\mathrm{bnd}}^\infty D(r)\ 2 \pi r dr = 
\frac{1}{2}\frac{GM_{\star}\dot{M}}{R_\mathrm{bnd}}.
\hfill{}
\label{eq: accretion energy}
\end{eqnarray} 

Lynden-Bell \& Pringle~(\cite{pringle}) assumed that the disk extends down to the stellar radius 
$R_\mathrm{bnd}=R_\mathrm{\star}$, or to the corotation radius where the Keplerian orbital period 
equals the stellar rotation period. Recent observations (e.\/g., Muzerolle et al.~\cite{muzerolle}) 
confirmed a magnetically mediated accretion processes in the innermost disk. The accretion disk is truncated 
by the stellar magnetic field at a radius $R_\mathrm{bnd}$ of several stellar radii 
and material is channeled along magnetic field lines onto the star with nearly free-fall velocity 
(e.\/g., Uchida \& Shibata~\cite{uchida}; Bertout et al.~\cite{bertoutII}). The magnetic flux density 
$B$ of a T\,Tauri star is in the range of $\sim$$1 - 3\ \mathrm{kG}$ (e.\/g., Johns-Krull et 
al.~\cite{johns-krull}). As the boundary radius $R_\mathrm{bnd}$ is assumed to be smaller than the 
sublimation radius of dust ($\approx$$R_\mathrm{in}$ for RY\,Tau), the accreting material in this inner region is gaseous. 
Close to the stellar surface the accreting, free-falling material is abruptly stopped in one or a 
series of shock layers. In this inner accretion zone the total released energy is 
$\zeta G M_{\star} \dot{M} / R_\mathrm{\star}$ with $\zeta = 1 - R_\mathrm{\star} / 
(2R_\mathrm{bnd})$. Therefore, adjacent layers of the shock region like the stellar photosphere are 
heated to temperatures of $T_\mathrm{bnd}=5700\ \mathrm{K}$ up to $ 8800\ \mathrm{K}$, while 
temperatures of $9000$ up to $20,000\ \mathrm{K}$ are achieved in upper shock areas (Calvet \& 
Gullbring~\cite{gullbring}). Muzerolle et al.~(\cite{muzerolle}) found in their studies that the 
energy contribution from the upper shock layers can be neglected for accretion rates $\dot{M} \lse 
10^{-6}\ \mathrm{M_\mathrm{\odot} yr^{-1}}$ while the stellar photosphere emits most of the accretion 
energy. Thus, the stellar photosphere is assumed to emit the intrinsic stellar flux {\it and} 
the accretion energy $\zeta G M_{\star} \dot{M} / R_\mathrm{\star}$. 

The crux of the accretion theory is the surface coverage factor $f$ which mimics an annulus on the 
stellar photosphere where the additional accretion emission occurs (D'Alessio et 
al.~\cite{dalessioIII}; Bertout et al.~\cite{bertoutII}; Lynden-Bell \&
Pringle~\cite{pringle}). If a blackbody emitter is assumed, one gets: 
\begin{eqnarray}
\hfill{}
f 4 \pi R_\mathrm{\star}^2 \sigma T_\mathrm{bnd}^4 = 
\zeta G M_{\star} \dot{M} / R_\mathrm{\star}.
\hfill{}
\label{blackbody}
\end{eqnarray}
The Stefan-Boltzmann constant is represented by $\sigma$. Calvet \& Gullbring~(\cite{gullbring}) 
determined $f \approx 0.1\%-1.0\%$ for HAeBe stars and $f \approx 10\%$ for T\,Tauri stars. 
Stronger veiling of absorption lines indicates the larger coverage factors $f$ of T\,Tauri stars.

\begin{figure*}[!tb]
\centering
\resizebox{\textwidth}{!}{\includegraphics[angle=-90]{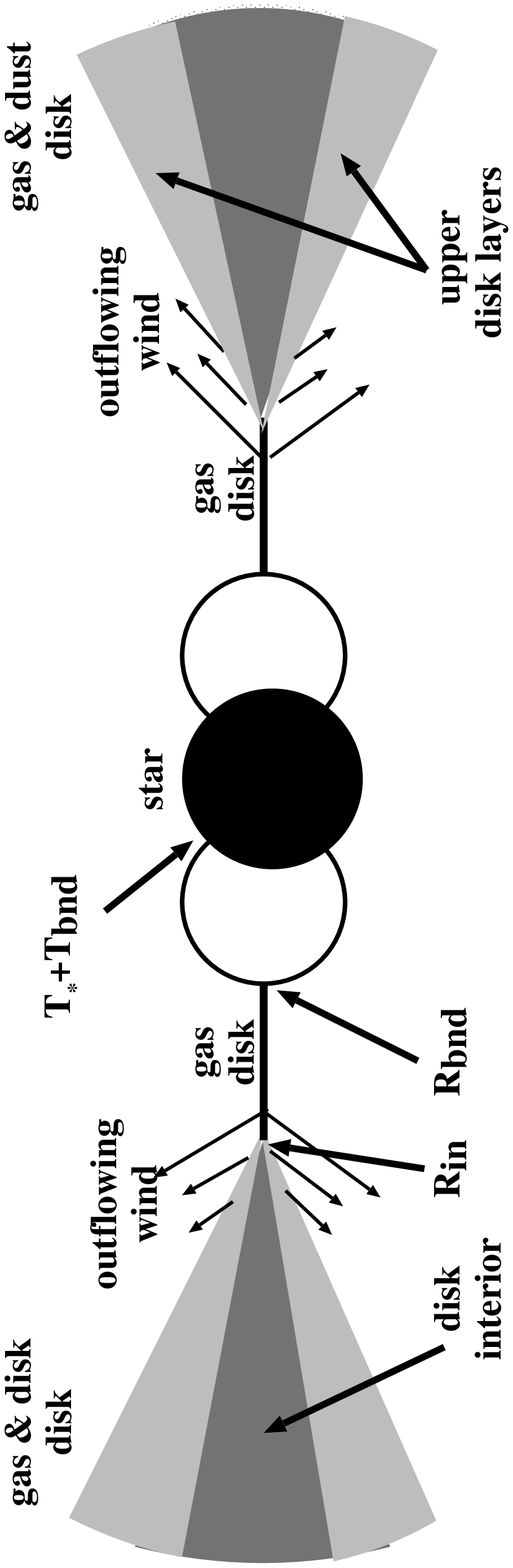}}
\caption{Simplified sketch of our active disk model. The boundary temperature 
$T_\mathrm{bnd}$, the boundary radius $R_\mathrm{bnd}$ and the inner disk radius $R_\mathrm{in}$ are 
shown. We use a two-layer disk model (Sect.~\ref{section:dust}): a disk atmosphere where the 
optical depth $\tau_\mathrm{N}$ is smaller than unity in N band  and a optical thick, disk interior.}
\label{figure:accretiondisk}
\end{figure*}

In the region between the sublimation and magnetic boundary radius the
gradually released accretion rate in our model 
(see Eq.~\ref{eq: dissipation rate}) is determined by the accretion theory of Lynden-Bell \& 
Pringles~(\cite{pringle}). Because of the steep decrease of the dissipation rate $D$ with increasing 
distance $r$ (see Eq.~\ref{eq: dissipation rate}) accretion at radial distances $r > R_\mathrm{in}$ 
is neglected in our model. The remaining gravitational energy of the 
accreting material between $r = R_\mathrm{bnd}$ and $r = R_\mathrm{\star}$ is emitted at the 
stellar surface assuming a blackbody emitter with the temperature $T_\mathrm{bnd}$. Note 
that the effects of the boundary temperatur $T_\mathrm{bnd}$ and the magnetic truncation 
radius $R_\mathrm{bnd}$ on the SED and MIR visibilities are marginal (see D'Alessio et 
al.~\cite{dalessioIII}). A sketch of our active disk model is shown in 
Fig.~\ref{figure:accretiondisk}. 

\end{document}